\documentclass[11pt, oneside]{article} 
\usepackage[margin=2cm]{geometry}                		
\usepackage[parfill]{parskip}    		
\usepackage{graphicx}			
\usepackage[utf8]{inputenc}
\usepackage[english]{babel} 
\usepackage{amsmath,amssymb,amsthm,bm, mathtools} 
\usepackage[margin=2cm]{geometry}
\usepackage[color=yellow]{todonotes}
\usepackage[colorlinks]{hyperref}
\usepackage{enumerate}
\usepackage{ulem}
\usepackage{tabularx, booktabs}

\usepackage{pgfplots}
\usepackage{float}
\usepackage{tikz,tikz-cd}
\usepackage{graphicx}
\usepackage{caption}
\usepackage{multicol}
\usepackage{subcaption}
\usepackage{natbib}
\extrafloats{100}
\numberwithin{equation}{section}

\newtheorem{theorem}{Theorem}[section]
\newtheorem{lemma}[theorem]{Lemma}

\newtheorem{proposition}[theorem]{Proposition}
\newtheorem{remark}[theorem]{Remark}

\theoremstyle{definition}

\newcommand{\ad}{\mathop{\mathrm{ad}}\nolimits}
\DeclarePairedDelimiter{\dangle}{\langle\!\langle}{\rangle\!\rangle}

\pgfplotsset{compat=1.16}

\def\contract{\makebox[1.2em][c]{\mbox{\rule{.6em}
{.01truein}\rule{.01truein}{.6em}}}}

\title{Irreversible dynamics for ideal fluids on Riemannian manifolds: metriplectic  energy-conserving alternatives to Navier--Stokes}
\author{Erwin Luesink, Archishman Saha, Tanya Schmah, Cristina Stoica}
\date{}

\begin{document}

\maketitle

\begin{abstract}
 For modelling dissipative phenomena such as friction, diffusion
and viscosity, the Hamiltonian framework has a key limitation:
once the Casimir invariants are fixed, Hamiltonian flows preserve
phase-space volume, which rules out relaxation to equilibrium. Thus
modelling dissipation requires modifying the Hamiltonian structure. 
Following the metriplectic (equivalently, GENERIC) approach,
we add to the Hamiltonian vector field a dissipative term generated by
a symmetric bracket, here constructed from a Riemannian metric. The resulting flow conserves the energy, and optionally further
invariants, while driving a chosen \textit{generalised entropy}
monotonically, which makes the dynamics \textit{irreversible}; when the entropy is bounded on energy level sets in the
direction in which it is driven, it acts as a Lyapunov functional. We
apply this construction to ideal fluids on compact Riemannian
manifolds, choosing Casimirs of the Lie--Poisson bracket as entropies. In
two dimensions, dissipating enstrophy yields the already-known energy-constrained
Navier--Stokes equations, which include a nonlocal term restoring energy balance. 
In three
dimensions, where enstrophy is no longer a Casimir, we derive new irreversible models. For
incompressible and compressible fluids carrying an advected scalar,
dissipating potential enstrophy or higher potential-vorticity moments
gives a nonlinear, anisotropic diffusion that acts to align the
gradients of the scalar and of the potential vorticity. For a purely
incompressible fluid, taking helicity as the entropy gives a
first-order, non-smoothing dissipative term whose stationary states are
Beltrami fields.   
\end{abstract}


\section{Introduction}

Hamiltonian mechanics on Poisson manifolds is a foundational framework
for modelling fluid flow \cite{shepherd1990symmetries, morrison1998hamiltonian} and many other physical systems. However, for
modelling \textit{dissipative} phenomena such as friction, diffusion
and viscosity, the Hamiltonian framework has a fundamental limitation:
on each symplectic leaf of phase space---for fluids, a set of isovortical flows, on which all Casimir invariants are fixed---the flow preserves
phase-space volume, so no open subset of a leaf can relax to an
equilibrium in that leaf; in particular, no equilibrium is
asymptotically stable within its leaf. Describing relaxation, mixing
and decay in real fluids therefore requires some modification of the
Hamiltonian structure.

Modern foundational approaches often rely on the concept of dissipativity as formalised in systems theory, a notion that traces back to the seminal work of \cite{willems1972dissipative}. Another modern geometric approach makes use of contact geometry, see for instance \cite{grmela2014contact, bravetti2019contact}. In contact Hamiltonian mechanics, one typically considers a symplectic Hamiltonian, $H_{cons}$, and augments it with a term that depends on the contact variable $z$ (interpreted as action or negative entropy), i.e., $H(q, p, z) = H_{cons}(q, p) + \alpha(z)$. Here, $\alpha$ acts as a friction function or damping function, often enforcing exponential contraction of phase space volume.

Another class of formulations relies on maintaining energy conservation while dissipating, decaying, or dispersing another quantity, often identified as entropy (or negative entropy, consistent with the second law of thermodynamics). Such dissipative Hamiltonian systems date back to \cite{kaufman1984dissipative, morrison1986paradigm} and form the core of \textit{metriplectic mechanics}, discussed in detail in Section \ref{sec:generaltheory}. This theory builds on the dissipative-bracket precursors of \cite{kaufmanmorrison1982algebraic, kaufman1984dissipative, morrison1984bracketPLA, morrison1984bracketsPAM, morrison1984observations, grmela1984particle, grmela1984bracket}. Parallel developments in \cite{grmela1997dynamics, ottinger1997dynamics} introduced the GENERIC (General Equation for Non-Equilibrium Reversible-Irreversible Coupling) framework. Axiomatically, metriplectic mechanics and GENERIC are equivalent. In both formulations, one starts with energy-preserving Hamiltonian mechanics on a Poisson manifold and introduces irreversible processes via postulating a symmetric bracket structure that respects the underlying geometry. The bracket viewpoint was developed further by Morrison and collaborators \cite{morrison1998hamiltonian, coquinot2020general} and, more recently, recast as a curvature-like metriplectic 4-bracket \cite{morrison2024inclusive, zaidni2025metriplectic}.
Closely related geometric constructions include the double-bracket and normal-metric gradient flows \cite{bloch1992completely, blochmorrisonratiu2013gradient}, metriplectic MHD \cite{materassi2012metriplectic}, and the generalized double-bracket vector fields on Poisson--Riemannian manifolds \cite{birtea2025generalized}. In \cite{mielke2025deriving}, the fundamental problem of coarse-graining infinite-dimensional Hamiltonian dynamics to obtain dissipative macroscopic dynamics is considered. The dissipative dynamics obtained has the GENERIC/metriplectic structure.

In the present paper, we introduce a constructive approach to irreversible fluid dynamics on Poisson manifolds that is closely related to the axiomatic approaches to irreversible mechanics. ``Irreversible'' here means that the time-reversal symmetry of Hamiltonian systems has been broken. 
It does not necessarily imply thermodynamic irreversibility. An alternative description is ``dissipative mechanics'',
where we mean dissipation of a Casimir instead rather than of energy. Fluid dynamics, particularly at large spatial and temporal scales, is well-modelled with geometric tools, see for instance \cite{shepherd1990symmetries}. In the general setting we do not insist that the entropy function is a Casimir, but merely require that it is a constant of motion that is functionally independent from the energy. Another difference is that we do not postulate a symmetric bracket, but rather we construct one explicitly by equipping the Poisson manifold with a Riemannian metric. We present the results on Riemannian manifolds as this allows us to treat different domains in a single framework. For geophysical fluid dynamics, one often works in spherical geometries, whereas for plasma physics, one often works with tori or twisted tori. These are compact domains, but they are curved, which plays an important role in the dissipative terms. By projecting the gradients of thermodynamic potentials onto subspaces orthogonal to the gradients of conserved quantities, we derive a class of dynamical systems that naturally exhibit selective decay. This geometric construction ensures that the first and second laws of thermodynamics are structurally built into the equations of motion. Our construction of the symmetric bracket is similar to the one in \cite{bloch2024metriplectic}. 

The specific examples we consider are two-dimensional incompressible fluids and three-dimensional compressible fluids on compact Riemannian manifolds. Morrison applied the metriplectic formalism to fluids, re-deriving the classical
Navier--Stokes--Fourier equations \cite{morrison1984bracketsPAM}. We work within the
same metriplectic structure but using a different symmetric bracket,
constructed geometrically from a Riemannian metric on the underlying Poisson
manifold. The resulting dynamics do not reproduce Navier--Stokes--Fourier; they
conserve energy while monotonically dissipating a chosen invariant, giving
constraint-respecting models of constrained-relaxation regimes, such as eddy transport,
plasma relaxation and geophysical mixing, where the classical equations are
intractable or too fine-grained. In particular, we show that the irreversible model built using the general theory leads to the Navier-Stokes equations in two dimensions with a nonlocal term to compensate for the energy loss if we pick the enstrophy as the entropy function. In three dimensions, applying the same framework cannot work directly, since the enstrophy is not a Casimir. By introducing an additional scalar field into the model, we obtain the Ertel potential vorticity as a scalar variable transported by the flow. From this quantity, it can be shown that there is an uncountably infinite family of integral invariants, among which one finds an enstrophy-like quantity. This quantity is the integral of the square of the Ertel potential vorticity and depends implicitly on the additional scalar variable. An application of the irreversible framework to this quantity yields a model that is different from the three-dimensional Navier-Stokes equations, because the induced dissipation is highly anisotropic. In view of the recent breakthrough with regards to the wellposedness problem of the three-dimensional Navier-Stokes equations with forcing, it is interesting to consider alternative models. We present such an alternative model that arises from a geometrically motivated framework.

The outline of the paper is as follows. In Section \ref{sec:generaltheory}, we present the general theory of irreversible mechanics on Poisson manifolds that are equipped with a Riemannian structure. In Section \ref{sec:twodimensions}, we discuss the Euler equations as geodesic equations on the group of volume-preserving diffeomorphisms, as introduced in the seminal paper \cite{arnold1966geometrie}. We then show how the irreversible mechanics framework gives rise to the Navier-Stokes equations with a nonlocal term to restore the energy balance on two-dimensional Riemannian manifolds and discuss generalisations of these equations. In Section \ref{sec:threedimensions}, we show that in the three-dimensional case, one does not obtain the Navier-Stokes equations upon following the same route as in the two-dimensional case, but instead one obtains a model with anisotropic nonlinear diffusion. 

\section{Metriplectic systems on ``Riemannian-Poisson'' manifolds}\label{sec:generaltheory}

Consider a manifold $M$ with both a Riemannian metric $\langle\,\cdot\,,\,\cdot\,\rangle$ and a Poisson bracket 
$\{\,\cdot\,,\,\cdot\,\}$.
We make no assumptions about compatibility of the two structures, e.g., we do not require that they form a Kähler or bi-Hamiltonian structure. (Note that some authors consider that the term ``Riemannian Poisson manifold'' implies some compatibility.)

We first summarise the general metriplectic paradigm of \cite{morrison1986paradigm}, which does \emph{not} require a Riemannian metric,
and then define the specific construction of a metriplectic bracket used in this paper.

\subsection{The metriplectic paradigm}

\cite{morrison1986paradigm} defines a class of dissipative systems called \textit{metriplectic systems}.
In its simplest form, the ingredients are: a Poisson manifold $(M, \{,\})$;
two scalar fields $\mathcal{H}$ (\textit{generalized energy}) and 
$\mathcal{S}$ (\textit{generalized entropy}),
and a positive semidefinite cometric with $d\mathcal{H}$ in its kernel.
The entropy $\mathcal{S}$ is required to be a conserved quantity, i.e. $\{\mathcal{H}, \mathcal{S}\} = 0$, and is 
usually chosen to be a \textit{Casimir}
of the Poisson bracket, i.e. $\{f, \mathcal{S}\} = 0$ for all $f \in C^\infty(M)$.
Morrison notes that in many physical 
contexts, including fluid dynamics, thermodynamic entropies are known to be Casimirs.
This justifies his focus on generalised entropies that are Casimirs, but the framework does not require this.

The cometric has an associated symmetric bracket on $\mathcal{C}^\infty$ defined by
\begin{equation}
    (f,g) \coloneqq\dangle{ d f, d g},
\end{equation}
with the degeneracy property 
\begin{align}\label{eq:symdegen1}
(\mathcal{H}, f) = 0 \quad \textrm{for all } f,
\end{align}
since $d\mathcal{H}$ is assumed to be in the kernel of the cometric.
Define the 
\textit{(generalised) free energy} $\mathcal{F}\in C^\infty(M)$ by 
\[
\mathcal{F}=\mathcal{H}-\mathcal{S},
\]
following the Helmholtz free-energy convention.
The corresponding dynamical system is defined by
\begin{align}
    \frac{d}{dt}f &= \{f,\mathcal{F}\} - \, (f,\mathcal{F})_\mathcal{H}, \qquad \textrm{(single-generator form)} \label{eq:singlegen1}\\
    &= \{f,\mathcal{H}\} - \{f,\mathcal{S}\} +  \, (f,\mathcal{S})_\mathcal{H}, \qquad \textrm{(two-generator form)} \label{eq:twogen1}
\end{align}
where the second line uses the bilinearity of both brackets and the degeneracy of the symmetric bracket
\eqref{eq:symdegen1}.
(These two forms are equivalent, for all $\mathcal{S}$.) 
Using the same degeneracy property, and the assumption $\{\mathcal{H},\mathcal{S}\}=0$, it follows that $\mathcal{H}$ is conserved:
\begin{equation}
    \begin{aligned}
    \frac{d}{dt}\mathcal{H} &=  \{\mathcal{H},\mathcal{H}\} - \{\mathcal{H},\mathcal{S}\} + (\mathcal{H},\mathcal{S}) =  0.
    \end{aligned}
\end{equation}
For the evolution of $\mathcal{S}$, using the same properties above, and also the skew-symmetry of the Poisson bracket, we obtain
\begin{equation}
    \begin{aligned}
        \frac{d}{dt}\mathcal{S} 
        &= \{\mathcal{S},\mathcal{H}\} - \{\mathcal{S},\mathcal{S}\} + (\mathcal{S},\mathcal{S}) 
        = (\mathcal{S},\mathcal{S}) \ge 0,
    \end{aligned}
\end{equation}
since the symmetric bracket is positive semidefinite.

\subsection{A metriplectic bracket for ``Riemannian-Poisson'' manifolds}\label{sec:RiemPoisson}

For any $\mathcal{K} \in C^\infty(M)$, define the projector $P^\perp_\mathcal{K}: TM \to TM$ by, for any
$v\in T_zM$,
\begin{equation}
    P^\perp_\mathcal{K}(v) = \begin{cases} v - \frac{\langle v,\nabla\mathcal{K}(z)\rangle}{\|\nabla\mathcal{K}(z)\|^2}\, \nabla\mathcal{K}(z), \quad &\|\nabla\mathcal{K}(z)\|\neq 0,\\
    v, &\text{otherwise},
    \end{cases}
\end{equation}
where $\nabla \mathcal{K}$ is the metric gradient of $\mathcal{K}$, 
defined by $\langle\nabla \mathcal{K}, X\rangle = d\mathcal{K}(X)$ for all vector fields $X$.
For any fixed $z \in M$ with $\nabla\mathcal{K}(z) \ne 0$, this is orthogonal projection onto the tangent to the level set of $K$ through $z$.
In case of a weak metric in the infinite-dimensional setting, the gradient can be replaced by a functional derivative,
as we do in Section \ref{sec:twodimensions}.  

We define the symmetric bracket $(\,\cdot\,,\,\cdot\,)_\mathcal{K}:C^\infty(M)\times C^\infty(M)\to C^\infty(M)$ by
\begin{equation}\label{eq:symK}
    (f,g)_\mathcal{K}\coloneqq\langle P^\perp_\mathcal{K}\nabla f, P^\perp_\mathcal{K}\nabla g\rangle =:\Sigma_\mathcal{K}(df,dg)
\end{equation}
Note that this bracket yields zero whenever either argument is a function of $\mathcal{K}$, and in particular,
\begin{align}\label{eq:symdegen}
(\mathcal{K}, f)_\mathcal{K} = (f,\mathcal{K})_\mathcal{K} = 0, \qquad \textrm{for any } f \in C^\infty(M) \qquad (\textit{degeneracy}).
\end{align}
By construction, the brackets $(\,\cdot\,,\,\cdot\,)_\mathcal{K}$ are bilinear and symmetric. 
From the tensorial form using $\Sigma_\mathcal{K}$, it is straightforward to deduce that the symmetric bracket satisfies the Leibniz property.

Let $\mathcal{H}\in C^\infty(M)$ and $\mathcal{S}\in C^\infty(M)$ be functionally independent and satisfy $\{\mathcal{H},\mathcal{S}\}=0$. In this context, we will call $\mathcal{H}$ the
\textit{(generalised) Hamiltonian} and $\mathcal{S}$ the \textit{(generalised) entropy}.
Note that $\mathcal{S}$ is an integral of motion of the Hamiltonian
system associated with $\mathcal{H}$. Define the 
\textit{(generalised) free energy} $\mathcal{F}\in C^\infty(M)$ by 
\[
\mathcal{F}=\mathcal{H}-\mathcal{S},
\]
following the Helmholtz free-energy convention.
Given a sign selector $\sigma = \pm 1$, we define the corresponding irreversible dynamics by
\begin{align}
    \frac{d}{dt}f &= \{f,\mathcal{F}\} - \sigma \, (f,\mathcal{F})_\mathcal{H}, \qquad \textrm{(single-generator form)} \label{eq:singlegen}\\
    &= \{f,\mathcal{H}\} - \{f,\mathcal{S}\} + \sigma \, (f,\mathcal{S})_\mathcal{H}, \qquad \textrm{(two-generator form)} \label{eq:twogen}
\end{align}
where the second line uses the bilinearity of both brackets and the degeneracy of the symmetric bracket
\eqref{eq:symdegen}.
(These two forms are equivalent, for all $\mathcal{S}$.) 
Using the same degeneracy property, and the assumption $\{\mathcal{H},\mathcal{S}\}=0$, it follows that $\mathcal{H}$ is conserved:
\begin{equation}
    \begin{aligned}
    \frac{d}{dt}\mathcal{H} &=  \{\mathcal{H},\mathcal{H}\} - \{\mathcal{H},\mathcal{S}\} + \sigma (\mathcal{H},\mathcal{S})_\mathcal{H} =  0.
    \end{aligned}
\end{equation}
For the evolution of $\mathcal{S}$, using the same properties above, and also the skew-symmetry of the Poisson bracket, we obtain
\begin{equation}
    \begin{aligned}
        \frac{d}{dt}\mathcal{S} 
        &= \{\mathcal{S},\mathcal{H}\} - \{\mathcal{S},\mathcal{S}\} + \sigma (\mathcal{S},\mathcal{S})_\mathcal{H}\\
        &= \sigma (\mathcal{S},\mathcal{S})_\mathcal{H}.
    \end{aligned}
\end{equation}
Since the symmetric bracket is positive semidefinite we see that, depending on the sign, $\mathcal{S}$ is either monotonically increasing or decreasing.

In the case $\sigma = 1$, the generalised entropy $\mathcal{S}$ is monotonically increasing, consistent with the behaviour of thermodynamic entropy.
However we will more often consider the case $\sigma = -1$, 
corresponding to a monotonic decreasing $\mathcal{S}$, a dynamical feature we refer to as
\textit{selective decay}, following \cite{gay2013selective}.
We call both systems \textit{irreversible}.

\begin{remark}
\cite{morrison1986paradigm} defines a similar, projection-based bracket in the free rigid body
example, using the implicit Euclidean metric on
$\mathfrak{so}(3)^*$ and an unnormalised orthogonal projection;
his bracket equals $\lambda \|\nabla\mathcal{H}\|^2$ times ours. 
\cite{bloch2024metriplectic} extend this to an arbitrary Riemannian metric on a
finite-dimensional Poisson manifold, defining
\[
K(df,dg) = \|\nabla\mathcal{H}\|^2\langle\nabla f,\nabla g\rangle
          - \langle\nabla\mathcal{H},\nabla f\rangle\langle\nabla\mathcal{H},\nabla g\rangle
          = \|\nabla\mathcal{H}\|^2\,(f,g)_\mathcal{H},
\]
where $(f,g)_\mathcal{H}$ is our bracket defined in \eqref{eq:symK}.
Morrison mentions that his framework allows a state-dependent prefactor $\lambda$,
so it potentially includes our normalisation, however this specific choice is not mentioned.
Our normalised bracket depends only on the level sets of $\mathcal{K}$:
it is unchanged under $\mathcal{K}\mapsto\varphi(\mathcal{K})$ with $\varphi'\neq 0$.
The price is an essential discontinuity of $(\,\cdot\,,\,\cdot\,)_\mathcal{K}$ at
critical points of $\mathcal{K}$. However, this does not affect the
dynamics in our examples. Indeed, the symmetric bracket enters \eqref{eq:singlegen} only
through $(f,\mathcal{S})_\mathcal{H}$, with
$|(f,\mathcal{S})_\mathcal{H}|\le\|\nabla f\|\,\|\nabla\mathcal{S}\|$, and in each
example of Sections~4--5, $\nabla\mathcal{S}$ vanishes at every critical point of
$\mathcal{H}$.
Moreover, when $\mathcal{H}$ is the kinetic energy its only critical point is
$u=0$, which, $\mathcal{H}$ being conserved, no trajectory with nonzero initial
energy approaches. Our choice of scaling leads to a interpretable ratio of competing physical quantities, see \eqref{E:nonlocal_ratio}.
\end{remark}

\begin{remark}
The symmetric bracket that \cite{morrison1984bracketsPAM} constructs
for fluids is of a very different character from the one used here. While ours depends
only on the Riemannian metric and the Hamiltonian, Morrison's is built from the phenomenological transport
coefficients $\eta$, $\zeta$ and $\kappa$ (shear viscosity, bulk viscosity, thermal conductivity), the thermodynamic state variables
$\rho$ and $T$, and the velocity gradient.
\end{remark}

If $\mathcal{S}$ is a Casimir, as will be the case below, then the only difference between our framework and Morrison's 
is the definition of the symmetric bracket.
Morrison's symmetric bracket may have zero eigenvalues because the underlying metric does, whereas in our framework the
underlying metric is nondegenerate but the symmetric bracket nonetheless has a zero eigenvalue in the $\nabla H$ direction.
However this difference is inessential. Morrison writes:
``Fundamentally it is desirable for dissipative systems to conserve dynamical constraints such as
energy and momentum while producing entropy.
For this reason one can build into the phase space
the preservation of such constants of motion. This
is done by requiring the phase space gradients of
these quantities to be null eigenvectors of [the metric].''
In particular, Morrison assumes that $\nabla \mathcal{H}$ is in the kernel of the metric. 
Given an underlying \textit{Riemannian} (i.e. nondegenerate) metric, a natural way to modify it so that $\nabla \mathcal{H}$ is in the kernel is 
to project vectors onto the orthogonal complement to $\nabla \mathcal{H}$ before applying the metric, as we have done in the previous section.
Thus our definition above is simply a constructive implementation of Morrison's metriplectic framework.

Note that, as discussed in \cite{morrison1986paradigm}, if $z_e$ is an isolated maximum of $\mathcal{S}$, 
then $\mathcal{S}$ is a strong Lyapunov function
and as a result $z_e$ is an asymptotically stable equilibrium point.

\begin{remark}
In the context of the metriplectic and GENERIC frameworks, $\mathcal{H}$ is typically the energy and $\mathcal{S}$ is an entropy functional which is also a Casimir of the Poisson bracket (i.e., $\{\mathcal{S},f\}=0$ for all $f$). In the present approach, the condition that $\mathcal{S}$ is a Casimir is not strictly required.
\end{remark}

\subsection{Multiple conserved quantities}
The formulation above preserves a single quantity $\mathcal{H}$. However, complex physical systems, particularly in fluid dynamics and plasma physics, often possess multiple Casimir invariants (e.g., enstrophy and its generalisations, magnetic helicity) that can be conserved by the irreversible dynamics simultaneously.

Let $\mathcal{I}_1, \dots, \mathcal{I}_k \in C^\infty(M)$ be a set of functionally independent conserved quantities we wish to maintain, 
with $\mathcal{I}_1 = \mathcal{H}$, and let $\mathcal{S}$ be the entropy we wish to dissipate 
(satisfying $\{\mathcal{S}, \mathcal{H}\}=0$, as before). We define the \textit{constrained symmetric bracket} by projecting gradients onto the orthogonal complement of the subspace spanned by $\{\nabla \mathcal{I}_1, \dots, \nabla \mathcal{I}_k\}$.

Let $\mathbf{G}$ be the Gram matrix of the gradients of the invariants, with entries $G_{ij} = \langle \nabla \mathcal{I}_i, \nabla \mathcal{I}_j \rangle$. Assuming independence, $\mathbf{G}$ is invertible. The multi-constraint projector $P^\perp_{\mathbf{I}}$ acting on a vector $v$ is
\begin{equation}
    P^\perp_{\mathbf{I}}(v) = v - \sum_{i,j=1}^k (G^{-1})_{ij} \langle v, \nabla \mathcal{I}_j \rangle \nabla \mathcal{I}_i.
\end{equation}
The dynamics are then governed by
\begin{equation}
    \frac{d}{dt} f = \{f, \mathcal{H}\} \pm \langle P^\perp_\mathbf{I} \nabla f, P^\perp_\mathbf{I} \nabla \mathcal{S} \rangle.
\end{equation}
By construction, $\frac{d}{dt}\mathcal{I}_k = 0$ for all $k$, while $\frac{d}{dt}\mathcal{S} \propto \pm|P^\perp{\mathbf{I}} \nabla \mathcal{S}|^2$.

\begin{remark}
The same multi-constraint bracket appears in dual form in \cite[Remark~2.1]{bloch2024metriplectic}.
\end{remark}

\section{Geometric framework for fluid dynamics}
\label{sec:geometric-framework}

The fluid systems considered in this paper share a common geometric
structure that we collect in this section before specialising to a particular
dimension or physical setting. Section~\ref{sec:twodimensions} will treat 2D incompressible flow, and then
Section~\ref{sec:threedimensions} will discuss 3D compressible and
incompressible flows. In the three-dimensional
case, the absence of a scalar vorticity necessitates the introduction
of an advected scalar field to produce a rich Casimir structure.
In each case the reversible dynamics admits a Lie--Poisson
Hamiltonian formulation on the dual of an appropriate Lie algebra, equipped
with a weak $L^2$ metric. For incompressible flow, with the metric identifying velocity fields with
their momenta, this dynamics is the Euler--Arnol'd formulation
\cite{arnold1966geometrie}; the Euler--Poincar\'e extension
\cite{holm1998euler} covers the case of advected quantities.
In all cases, the Lie algebra encodes the kinematics: volume-preserving vector fields for incompressible flow, a
semidirect product with $C^\infty(M)$ and densities when advected
quantities are present. The geometric scaffolding is otherwise
identical. This uniformity is what allows the symmetric-bracket
construction of Section~\ref{sec:RiemPoisson} to be applied
across the different fluid models without modification: in each case the
Casimirs of the Lie--Poisson bracket are the quantities conserved by the
reversible flow that will be dissipated by the irreversible dynamics.

Let $(M, \langle \cdot, \cdot \rangle)$ be a compact, orientable Riemannian
manifold without boundary, with Riemannian volume form $\mu$. The metric
induces the musical isomorphisms $\flat : \mathfrak{X}(M) \to \Omega^1(M)$
and $\sharp = \flat^{-1} : \Omega^1(M) \to \mathfrak{X}(M)$, defined by
$\langle u^\flat, X \rangle = \langle u, X \rangle$ for all vector fields
$X$. The Riemannian divergence of a vector field $u \in \mathfrak{X}(M)$ is
characterised by $\mathcal{L}_u \mu = (\mathrm{div}_\mu u) \mu$, and the
codifferential $d^*$ on $1$-forms satisfies $-d^* u^\flat = \mathrm{div}_\mu
u$. We write $\mathfrak{X}_{\mathrm{div}}(M)$ for the space of
divergence-free vector fields.

On the space of vector fields we use the weak Riemannian metric
\begin{equation}
\label{eq:weak-metric}
\langle u, v \rangle_{\mathfrak{X}}
= \int_M \langle u(x), v(x) \rangle\, \mu(dx),
\qquad u, v \in \mathfrak{X}(M),
\end{equation}
which induces an $L^2$-type topology rather than a smooth one. Under this
metric, elements of $\mathfrak{X}^*(M)$ are identified with covector-valued
densities: an element $m \in \mathfrak{X}^*(M)$ can be written as $m =
u^\flat \mu$ for some $u \in \mathfrak{X}(M)$, and the pairing of $m$ with
$v \in \mathfrak{X}(M)$ recovers the inner product in
\eqref{eq:weak-metric}. The same construction restricts to
$\mathfrak{X}_{\mathrm{div}}(M)$ and will be applied to both the
incompressible and the compressible settings.

Now if $F\in C^{\infty}(\mathfrak{X}^*_{\mathrm{div}}(M))$ is a smooth function, then its first variation at $m\in\mathfrak{X}^*_{\mathrm{div}}(M)$ defines a continuous linear functional on $\mathfrak{X}^*_{\mathrm{div}}(M)$, that is, an element of $\mathfrak{X}^{**}_{\mathrm{div}}(M)$. By means of the weak Riemannian metric, we can identify $\mathfrak{X}^{**}_{\mathrm{div}}(M)\simeq\mathfrak{X}_{\mathrm{div}}(M)$. The
G\^ateaux derivative at $m \in \mathfrak{X}^*(M)$ in the direction $\phi
\in \mathfrak{X}^*(M)$ is
\begin{equation}
\delta F[m; \phi]
= \lim_{\varepsilon \to 0}
\frac{F[m + \varepsilon \phi] - F[m]}{\varepsilon}.
\end{equation}
By the Riesz representation theorem associated with the weak metric
\eqref{eq:weak-metric}, there exists a unique element $\delta F/\delta m
\in \mathfrak{X}(M)$ such that
\begin{equation}
\label{eq:functional-derivative}
\delta F[m; \phi]
= \Bigl\langle \phi, \frac{\delta F}{\delta m}
\Bigr\rangle_{\mathfrak{X}},
\end{equation}
and we call $\delta F/\delta m$ the functional derivative of $F$ at $m$.
Throughout the paper we identify the dual space $\mathfrak{X}^*(M)$ with
$\mathfrak{X}(M)$ via the weak metric, and we use the same identification
for any Lie subalgebra of $\mathfrak{X}(M)$ or any extension by spaces of
advected quantities.

Given a Lie algebra $\mathfrak{g}$ that is a subspace or extension of
$\mathfrak{X}(M)$, equipped with the induced weak $L^2$ pairing, the
Lie--Poisson bracket on $C^\infty(\mathfrak{g}^*)$ is
\begin{equation}
\label{eq:LP-abstract}
\{F, G\}_{\mathfrak{g}}(m)
= \Bigl\langle m, \Bigl[ \frac{\delta F}{\delta m},
\frac{\delta G}{\delta m} \Bigr]_{\mathfrak{g}}
\Bigr\rangle,
\qquad m \in \mathfrak{g}^*.
\end{equation}
Here $[\cdot,\cdot]_{\mathfrak{g}} = -[\cdot,\cdot]_{JL}$, where
$[X,Y]_{JL} = \mathcal{L}_X Y = (X\cdot\nabla)Y - (Y\cdot\nabla)X$
is the Jacobi--Lie bracket of vector fields; we write
$\ad_\xi\eta = [\xi,\eta]_{\mathfrak{g}}$, and $\ad^*_\xi$ denotes the plain transpose of
$\ad_\xi$, $\langle \ad^*_\xi m, \eta\rangle = \langle m, \ad_\xi \eta\rangle$.
With these conventions $\ad^*_u m = \mathcal{L}_u m$ for $m\in\mathfrak{X}^*(M)$,
as in \cite[(6.2)]{holm1998euler}.
Equation \eqref{eq:LP-abstract} is the right (plus) Lie--Poisson
bracket appropriate to right-invariant systems, and Hamilton's equation
$\dot F = \{F, H\}_{\mathfrak{g}}$ generated by a Hamiltonian $H$ gives the
Lie-advection form
\begin{equation}
\label{eq:LP-eom}
\partial_t m + \mathcal{L}_{\delta H/\delta m}\, m = 0
\end{equation}
(modulo the constraints discussed below), the geometric form of the Euler
equations; this is the reduction convention of \cite{holm1998euler} for the
right-invariant (particle-relabelling) symmetry of a fluid. Specialising
$\mathfrak{g}$ to the relevant Lie algebra produces the Hamiltonian
formulation appropriate to each physical setup. For incompressible flow we
will take $\mathfrak{g} = \mathfrak{X}_{\mathrm{div}}(M)$ with the
bracket 
$[\cdot,\cdot]_{\mathfrak{g}} = -[\cdot,\cdot]_{JL}$ as above,
treated in
Section~\ref{sec:twodimensions} (in two dimensions) and in
Section~\ref{sec:threedimensions} (in three dimensions, where an
advected scalar is included); with $H$ the kinetic energy
\eqref{eq:weak-metric}, equation \eqref{eq:LP-eom} is the Euler--Arnol'd
equation \cite{arnold1966geometrie}, realising ideal incompressible flow as
geodesic motion on the volume-preserving diffeomorphism group. For
compressible flow with advected mass density and an advected scalar field,
we will take the semidirect product Lie algebra
\begin{equation}
\label{eq:semidirect-algebra}
\mathfrak{S}
= \mathfrak{X}(M) \ltimes \bigl( C^\infty(M) \times
\mathrm{Dens}_\mu(M) \bigr),
\end{equation}
in which the action of $\mathfrak{X}(M)$ on the scalar and density factors
is by Lie derivative. The corresponding Lie--Poisson bracket then acquires
additional terms reflecting the advection of the scalar and density fields,
and the equations of motion are of Euler--Poincar\'e type
\cite{holm1998euler}, as detailed in Section~\ref{sec:threedimensions}. In
contrast to the purely kinetic incompressible Hamiltonian, the compressible
Hamiltonian carries an internal-energy contribution, so the motion is no
longer geodesic.

A standard device, used in both the two- and three-dimensional settings
below, is to work with the full Lie algebra $\mathfrak{X}(M)$ and impose
incompressibility via a Lagrange multiplier $p \in C^\infty(M)$ enforcing
$d^* u^\flat = 0$. The multiplier $p$ is identified with the fluid
pressure, and the projection from $\mathfrak{X}(M)$ onto
$\mathfrak{X}_{\mathrm{div}}(M)$ is realised by the Leray projector,
whose explicit form depends on the Hodge decomposition of $1$-forms on
$M$ and is given in Sections~\ref{sec:twodimensions}
and~\ref{sec:threedimensions}.

The two roles of pressure in incompressible and compressible flow are
related by a singular limit. In the compressible setting, pressure is a
thermodynamic quantity determined by the internal energy and density, and
its gradient appears as a force in the momentum equation. In the
incompressible setting, pressure instead enters as the Lagrange multiplier
$p$ enforcing $d^* u^\flat = 0$. The incompressible equations may be
obtained as the limit in which the speed of sound becomes infinite:
density variations are suppressed energetically, and the pressure ceases
to derive from a thermodynamic relation, instead reacting instantaneously
to enforce the incompressibility constraint. This perspective places the
two settings within a single geometric framework, with the choice of Lie
algebra and Hamiltonian distinguishing them but the Lie--Poisson structure
common to both.

Together with the symmetric-bracket construction introduced in
Section~\ref{sec:RiemPoisson}, the data
$(\mathfrak{g}, \{\cdot, \cdot\}_{\mathfrak{g}},
\langle \cdot, \cdot \rangle_{\mathfrak{X}})$ make $\mathfrak{g}^*$ into a
Riemannian--Poisson manifold (in the weak, infinite-dimensional sense),
which is the setting in which the irreversible dynamics will be
constructed.

\section{Irreversible ideal fluid dynamics in 2D}\label{sec:twodimensions}
In this section, we describe irreversible formulations of ideal fluid dynamics on two-dimensional Riemannian manifolds. We start by describing the Hamiltonian formulation of ideal fluid dynamics, closely following \cite{arnold1966geometrie}, but specialising to the two-dimensional case.

\subsection{Reversible ideal fluids}
Let $(M,\langle\,\cdot\,,\,\cdot\,\rangle)$ be a two-dimensional compact, connected Riemannian manifold without boundary with Riemannian volume form $\mu$.
The Hodge star operator $\star:\Omega^k(M)\to \Omega^{2-k}(M)$ is the linear map defined by
\[
\alpha \wedge \star \beta = \langle \alpha, \beta \rangle\,\mu,
\]
for all $\alpha,\beta \in \Omega^k(M)$, where $\langle \cdot,\cdot\rangle$ denotes the pointwise inner product induced by the metric.
In particular, the Hodge star sends any function $f\in \Omega^0(M)$ to the top-form $f\,\mu$, and one has $\star^2=\mathrm{Id}$ on $\Omega^0(M)$ (equivalently also on $\Omega^2(M)$), while on $1$-forms the Hodge star acts as an endomorphism $\star:\Omega^1(M)\to\Omega^1(M)$ satisfying $\star^2=-\mathrm{Id}$. 

The space of volume-preserving diffeomorphisms is defined as
\begin{equation}
\mathrm{Diff}_{\mu}(M)=\{\varphi\in \mathrm{Diff}(M)\,| \,  \varphi^*\mu = \mu\}.
\end{equation} 
On a two-dimensional Riemannian manifold, the Riemannian volume form is a symplectic form, i.e., a nondegenerate closed two-form. Consequently, the group $\mathrm{Diff}_\mu(M)$ coincides with the group of symplectomorphisms of $(M,\mu)$. Infinitesimally, if $X\in\mathfrak{X}(M)$ denotes the vector field generating a differentiable one-parameter family $t\in \mathbb{R}\mapsto\varphi_t\in\mathrm{Diff}(M)$, the condition $\varphi_t^*\mu = \mu$ differentiates to 
\begin{equation}
    \mathcal{L}_X\mu = 0,
\end{equation}
which, using Cartan's identity $\mathcal{L}_X = \iota_X d+ d\iota_X$ implies $d(\iota_X\mu)=0$. It follows that $X$ is a divergence-free vector field with respect to $\mu$
\begin{equation}
    \mathfrak{X}_{\mathrm{div}}(M) = \{X\in \mathfrak{X}(M)\,|\, \mathrm{div}_\mu X = 0\},
\end{equation}
and the Lie algebra of $\mathrm{Diff}_\mu(M)$ coincides with the space of symplectic (equivalently, divergence-free) vector fields on $(M,\mu)$, equipped with minus the Jacobi-Lie bracket of vector fields. If $M$ is simply connected, every symplectic vector field is Hamiltonian, which allows us to establish an anti-homomorphism between $(\mathfrak{X}_\mathrm{div}(M),[\,\cdot\,,\,\cdot\,])$ and $(C^\infty(M)/\mathbb{R},\{\,\cdot\,,\,\cdot\,\}_{\mathrm{can}})$, where $\{\,\cdot\,,\,\cdot\,\}_{\mathrm{can}}$ is the canonical Poisson bracket. If $M$ is not simply connected, then we appropriately adapt the function space to absorb the arising harmonic forms, see \cite[p. 311]{marsden1983coadjoint}. For the torus $\mathbb{T}^2$, it is necessary to restrict to flows with zero mean velocity. The quotient by constants is necessary because constant functions generate the zero Hamiltonian vector field. 

We may identify the dual space $\mathfrak{X}^*_{\mathrm{div}}(M)$ with $\mathfrak{X}_{\mathrm{div}}(M)$ under the weak Riemannian metric. Under this identification, the Lie-Poisson bracket induces a Poisson structure directly on $\mathfrak{X}_{\mathrm{div}}(M)$. Combining all of the results so far, we have established that $(\mathfrak{X}_{\mathrm{div}}(M), \{\,\cdot\,,\,\cdot\,\}_{\mathfrak{X}}, \langle\,\cdot\,,\,\cdot\,\rangle_\mathfrak{X})$ is a Riemannian-Poisson manifold. We define the Hamiltonian in terms of $u$, or equivalently $u^\flat$, using the metric or the cometric as follows, \begin{equation}\label{eq:hameuler}
    \mathcal{H}(u) = \frac{1}{2}\int_M\langle u,u\rangle \mu = \frac{1}{2}\int_M u^\flat\wedge\star u^\flat = \frac{1}{2}\langle u,u\rangle_\mathfrak{X} = \frac{1}{2}\|u^\flat\|^2_{L^2} = \mathcal{H}(u^\flat)
\end{equation}
which is simply the kinetic energy of the incompressible flow. With the definition of the functional derivative as above, we compute $\frac{\delta \mathcal{H}}{\delta u} = u$ and $\frac{\delta\mathcal{H}}{\delta u^\flat} = u^\flat$. Let us consider now the vector field formulation. Via integration by parts, we obtain the Lie-Poisson dynamics formulated in terms of the coadjoint action corresponding to a right action of the diffeomorphism group
\begin{equation}
\frac{\partial}{\partial t}u + \left(\mathrm{ad}^*_u u^\flat\right)^\sharp = 0, \quad u\in\mathfrak{X}_{\mathrm{div}}(M).
\end{equation}
This is the Euler-Arnold equation for ideal inviscid fluid dynamics, which reveals that ideal fluid dynamics can be viewed as geodesic motion on the group of volume-preserving diffeomorphisms for a right-invariant metric. To relate Euler-Arnold equation to the usual presentation of the equations of motion in fluid dynamics, it is convenient to express the dynamics in the full Lie algebra of vector fields $\mathfrak{X}(M)$ rather than just the divergence-free ones. The coadjoint action does not automatically stay divergence-free, so it must be projected. One then writes
\begin{equation}
    \frac{\partial}{\partial t}u + \mathbb{P}\left(\mathrm{ad}^*_u u^\flat\right)^\sharp\ = 0, \quad u\in\mathfrak{X}(M),
\end{equation}
where $\mathbb{P}:\mathfrak{X}(M)\to\mathfrak{X}_{\mathrm{div}}(M)$ is the Leray projector, for which an elegant closed form expression can be given. We can express the Riemannian divergence of a vector field $v\in\mathfrak{X}(M)$ as $-d^*v^\flat$, where $d^* = -\star d\star: \Omega^{k}(M) \rightarrow \Omega^{k - 1}(M)$ is the codifferential for 1-forms. The Hodge decomposition tells us that any 1-form $v^\flat$ for $v\in\mathfrak{X}(M)$ can be uniquely expressed as
\begin{equation}
    v^\flat = d\chi + \star d\psi,
\end{equation}
where $\chi\in C^\infty(M)$ is called the velocity potential and $\psi\in C^\infty(M)$ is the stream function. Since $M$ is simply connected, compact, and has no boundary, the only harmonic 1-form is the zero form. Computing the Riemannian divergence of $v$ yields
\begin{equation}
    -d^*v^\flat = -d^*d\chi - d^*\star d\psi = -d^*d\chi + \star d\star\star d\psi = -d^*d\chi - \star d^2\psi = -d^*d\chi,
\end{equation}
since $\star\star=-1$ for 1-forms and $d^2 = 0$. Hence only $d\chi$ contributes to the divergence and $\star d\psi$ is always divergence-free. It follows that the Leray projector is explicitly given by
\begin{equation}\label{eq:Leray}
    \mathbb{P}v = v-(d\chi)^\sharp.
\end{equation}
We can thus write the Euler equations for ideal fluids on a two-dimensional compact simply
connected Riemannian manifold without boundary in terms of vector fields and in terms of
1-forms as
\begin{equation}\label{eq:euler}
\begin{aligned}
\frac{\partial u}{\partial t} + \left(\operatorname{ad}^*_u u^\flat\right)^\sharp &= -(dp)^\sharp,
&\qquad
\frac{\partial u^\flat}{\partial t} + \mathcal{L}_u u^\flat &= -dp,\\[4pt]
\operatorname{div}_\mu u &= 0,
&\qquad
d^* u^\flat &= 0.
\end{aligned}
\end{equation}
Here the pressure $p\in C^\infty(M)$ serves as the Lagrange multiplier enforcing
incompressibility. The left formulation is the Euler–Arnold equation, making explicit the
geodesic nature of the flow on $\mathfrak{X}_{\mathrm{div}}(M)$, while the right formulation
is the Euler–Poincaré equation governing the evolution of the momentum 1-form
$u^\flat$. We refer to \cite{holm1998euler} for a extensive treatment of Euler-Poincar\'e equations in continuum mechanics. For divergence-free $u$, one has the identity
$\operatorname{ad}^*_u u^\flat = \mathcal{L}_u u^\flat$, but writing $\mathcal{L}_u u$ would
be incorrect (since $\mathcal{L}_u u = 0$). Thus we retain the $\operatorname{ad}^*$ notation
in the Euler–Arnold formulation to emphasize the coadjoint action.

\begin{remark}
    Fluids with advected quantities are usually modelled on semidirect products of the diffeomorphism group with spaces of tensors over compact Riemannian manifolds. The Euler-Poincar\'e description continues to exist, but the Euler-Arnold formulation is often lost. The reason is that Hamiltonians consisting of kinetic and potential energy are not necessarily Riemannian metrics.
\end{remark}

By means of the homomorphism between $(\mathfrak{X}_{\mathrm{div}}(M),[\,\cdot\,,\,\cdot\,])$ and $(C^\infty(M)/\mathbb{R}, \{\,\cdot\,,\,\cdot\,\}_{\mathrm{can}})$, we can formulate the equations of ideal fluid motion in terms of vorticity and stream function. The vorticity 2-form $\star\omega$ is obtained by exterior differentiating the velocity 1-form $u^\flat$
\begin{equation}
    \star\omega = du^\flat = d\star d\psi,
\end{equation}
where we recall that $u^{\flat} = \star d\psi$ since $u$ is divergence-free. Applying the Hodge star to the equation above and recalling that codifferential of $1$-forms satisfies $d^*=-\star d\star$, we obtain the scalar vorticity,
\begin{equation}\label{eq:starduflat}
    \omega = \star d\star d\psi = -d^*d\psi= \Delta\psi,
\end{equation}
where $\Delta$ is the Laplace-Beltrami operator and $dd^*\psi=0$ since $\psi$ is a function. We have adopted here the geometer's sign convention $\Delta= -(dd^* + d^*d)$, which means that the Laplace-Beltrami operator is negative-definite. This convention means that the Hodge- De Rham Laplacian in Euclidean coordinates coincides with the usual Laplacian in Euclidean space. The direct relation between the velocity 1-form $u^\flat$ and the scalar vorticity is thus $\omega = \star du^\flat$. Applying $\star d$ to the 1-form formulation in \eqref{eq:euler} yields the vorticity-stream function formulation of fluid dynamics
\begin{equation}
    \begin{aligned}
    \frac{\partial}{\partial t}\omega+\{\psi,\omega\}_{\mathrm{can}} &= 0,\\
    \omega &= \Delta\psi,
    \end{aligned}
\end{equation}
where $\{\,\cdot\,,\,\cdot\,\}_{\mathrm{can}}$ is the canonical Poisson bracket on $M$ (which is a symplectic manifold because volume-preserving diffeomorphisms are symplectomorphisms for two-dimensional $M$). That is $\{f,g\}_{\mathrm{can}}\mu = df\wedge dg$, where $\mu$ is the volume form on $M$. The Hamiltonian \eqref{eq:hameuler} can be formulated in terms of vorticity as well. Since $u^\flat = \star d\psi$ for incompressible flows, we have
\begin{equation}
    \frac{1}{2}\int_M\langle u,u\rangle\mu = \frac{1}{2}\int_M u^\flat\wedge\star u^\flat = \frac{1}{2}\int_M d\psi\wedge\star d\psi = -\frac{1}{2}\int_M \psi\Delta\psi\mu = -\frac{1}{2}\int_M\omega\Delta^{-1}\omega\mu,
\end{equation}
by using integration by parts. With the conventions that we have taken, the Laplace-Beltrami operator is negative semidefinite so the Hamiltonian remains a nonnegative function. The Hamiltonian can thus be formulated in terms of vorticity as
\begin{equation}
    \mathcal{H}(\omega) = -\frac{1}{2}\int_M\omega\Delta^{-1}\omega\mu = -\frac{1}{2}\int_M\star\omega\Delta^{-1}\omega.
\end{equation}
Note that the Hamiltonian pairs scalar functions and volume forms. This enables point vortex descriptions for two-dimensional fluids. The variational derivative of the Hamiltonian with respect to $\omega$ is given by
\begin{equation}
    \frac{\delta \mathcal{H}}{\delta \omega} = -\Delta^{-1}\omega = -\psi,
\end{equation}
and hence the equations of ideal fluid dynamics on a two-dimensional orientable simply connected compact Riemannian manifold are given in Hamiltonian form by
\begin{equation}
    \frac{\partial}{\partial t}\omega - \Big\{\frac{\delta \mathcal{H}}{\delta\omega},\omega\Big\}_{\mathrm{can}} = 0.
\end{equation}
 The Lie-Poisson bracket introduced in \eqref{eq:LP-abstract} can also be transformed to the vorticity formulation. In terms of the vorticity, the bracket is given by
\begin{equation}
    \{F,G\}_{\mathfrak{X}}(\omega) = \left\langle \omega,\Big\{\frac{\delta F}{\delta\omega},\frac{\delta G}{\delta\omega}\Big\}_{\mathrm{can}}\right\rangle_\mathfrak{X} 
    = \int_M \omega(x)\Big\{\frac{\delta F}{\delta\omega},\frac{\delta G}{\delta\omega}\Big\}_{\mathrm{can}}(x)\mu(dx).
\end{equation}
It is now straightforward to show that the Lie-Poisson bracket $\{F,C\}=0$ for all $F\in C^\infty(M)$ if $C$ is given by
\begin{equation}
C = \int_M\Phi(\omega)\mu,    
\end{equation}
where $\Phi:C^\infty(M)\to\mathbb{R}$ is any smooth analytic function. These invariants are the Casimirs for ideal fluid dynamics. While analyticity of $\Phi$ is not strictly necessary for proving that these quantities do not evolve in time, it does allow us to connect these Casimirs to the moment hierarchy of fluid invariants. Indeed, if $\Phi$ is analytic, then it has a convergent Taylor expansion
\begin{equation}
\Phi(\omega) = \sum_{n=0}^\infty a_n\omega^n,    
\end{equation}
and thus
\begin{equation}
    C = \sum_{n=0}^\infty a_n\int_M \omega^n \mu.
\end{equation}
Several of the lowest order moments correspond to important physical invariants. The zeroth moment is simply the total volume of the domain $M$ and is trivially constant, the first moment expresses conservation of total vorticity and the second moment is the enstrophy, which plays an important in the theory of two-dimensional turbulence.

\subsection{Irreversible ideal fluids}
In the previous section we established that the space $(\mathfrak{X}_\mathrm{div}(M),\{\,\cdot\,,\,\cdot\,\}_\mathfrak{X},\langle\,\cdot\,,\,\cdot\,\rangle_\mathfrak{X})$ is a Riemannian-Poisson manifold. We discussed the reversible motion that can be constructed from it and showed that there is an infinite family of constants of motion. In this section, we employ the theory developed in Section \ref{sec:generaltheory} to introduce a symmetric bracket defined in terms of variational derivatives.

We first formulate the symmetric bracket in terms of the velocity field. Let the first constant of motion be the Hamiltonian $\mathcal{H}(u)$ as given in \eqref{eq:hameuler}. Any non-constant Casimir Poisson-commutes with the Hamiltonian and is functionally independent of it. Hence any Casimir is a valid choice for the function $\mathcal{S}$:
\begin{equation}
    \mathcal{S}(u) = \int_M\Phi(\star du^\flat)\mu =\int_M\Phi(\omega)\mu,
\end{equation}
for some smooth analytic function $\Phi$. Given the Riemannian metric $\langle\,\cdot\,,\,\cdot\,\rangle_\mathfrak{X}$, we define the symmetric bracket
\begin{equation}\label{eq:symbrackfluids}
    (F,G) = \left\langle P^\perp_\mathcal{H}\frac{\delta F}{\delta u}, P^\perp_\mathcal{H}\frac{\delta G}{\delta u}\right\rangle_\mathfrak{X},
\end{equation}
where the projector is given by
\begin{equation}
    P^\perp_\mathcal{H}(v) = \begin{cases}
        v - \dfrac{\langle v,\frac{\delta\mathcal{H}}{\delta u}\rangle_\mathfrak{X}}{\|\frac{\delta\mathcal{H}}{\delta u}\|_\mathfrak{X}^2}\dfrac{\delta\mathcal{H}}{\delta u},\quad &\Big\|\dfrac{\delta\mathcal{H}}{\delta u}\Big\|_\mathfrak{X}\neq 0,\\[8pt]
        v,& \text{otherwise.}
    \end{cases}
\end{equation}
We now compute using the Gateaux derivative, integration by parts and the fact that $\star\star=-1$ for 1-forms in two dimensions that the first variation of a general Casimir is given by
\begin{equation}
    \delta\mathcal{S} = \int_M \Phi'(\omega)\delta\omega \,\mu = \int_M\Phi'(\omega) (\star d \delta u^\flat) \mu = -\int_M\langle\star d\Phi'(\omega),\delta u^\flat\rangle\mu = -\langle (\star d\Phi'(\omega))^\sharp,\delta u\rangle_\mathfrak{X} .
\end{equation}
Hence the variational derivative is given by
\begin{equation}
    \frac{\delta\mathcal{S}}{\delta u} = -(\star d\Phi'(\omega))^\sharp.
\end{equation}
It follows directly from this computation that the total vorticity, given by $\mathcal{S}_1(\omega)=\int_M\omega\mu$, has a vanishing variational derivative with respect to $u$, since $\Phi'(\omega)=1$ in that case so $d\Phi'(\omega)=0$, and would therefore induce a trivial symmetric bracket, and thereby a trivial irreversible term. 

\subsubsection{Enstrophy} \label{sssect:enstrophy}
The simplest nontrivial choice is the enstrophy, given by $\mathcal{S}_2(\omega)=\frac{1}{2\,\mathrm{Re}}\int_M\omega^2\mu$, where we have chosen to introduce the Reynolds number $\mathrm{Re}=\frac{UL}{\nu}$. The Reynolds number relates the typical velocity scale $U$ and typical length scale $L$ to the kinematic viscosity $\nu$. For the enstrophy, the variational derivative is given by 
\begin{equation}\label{eq:varderivativeenstrophy}
    \frac{\delta\mathcal{S}_2}{\delta u} = -\frac{1}{\mathrm{Re}}(\star d\omega)^\sharp.
\end{equation}
The projection of the variational derivative of the enstrophy onto the orthogonal direction of the variational derivative $\frac{\delta\mathcal{H}}{\delta u} =u$ for the Hamiltonian given in \eqref{eq:hameuler} is 
\begin{equation}\label{eq:projenstrophy}
    P^\perp_\mathcal{H}\frac{\delta \mathcal{S}_2}{\delta u} = -\frac{1}{\mathrm{Re}}(\star d\omega)^\sharp + \frac{\langle(\star d\omega)^\sharp, u\rangle_\mathfrak{X}}{\mathrm{Re}\,\|u\|^2_\mathfrak{X}}u = -\frac{1}{\mathrm{Re}}\Big((\star d\omega)^\sharp + c(u)u\Big)
\end{equation}
if $\|u\|^2_\mathfrak{X}\neq 0$. The scalar function $c(u)$ measures how the domain-averaged vorticity changes in the cross-stream direction and normalises this by the total kinetic energy. From \eqref{eq:projenstrophy} it can be seen that $c(u)$ is explicitly given by
\begin{equation}\label{E:nonlocal_ratio}
    c(u) = -\frac{\langle(\star d\omega)^\sharp, u\rangle_\mathfrak{X}}{\|u\|^2_\mathfrak{X}} =  \frac{\int_M\omega(\star du^\flat)\mu}{\|u\|^2_\mathfrak{X}} = \frac{\|\omega\|^2_{L^2}}{\|u\|^2_\mathfrak{X}},
\end{equation}
so another interpretation of $c(u)$ is the ratio of enstrophy to energy. Note that the $L^2$-norm of the velocity field uses the Riemannian metric and both norms are defined with respect to the Riemannian volume form. Since $c(u)$ depends on norms, it corresponds to a nonlocal term in the equations of motion. Let the free energy $\mathcal{F} = \mathcal{H} - \mathcal{S}_2$. Then the enstrophy-minimising irreversible dynamics of any observable $f\in C^\infty(\mathfrak{X}_\mathrm{div}(M))$ is given by
\begin{equation}\label{eq:enstrophy_minimising_bracket}
    \frac{d}{dt}f = \{f,\mathcal{H}\} - (f,\mathcal{S}_2).
\end{equation}
It is then immediate from the definition of the symmetric bracket in \eqref{eq:symbrackfluids} that, for sufficiently regular solutions, $\frac{d}{dt} \mathcal{H} = 0$ and $\frac{d}{dt}\mathcal{S}_2 = -(\mathcal{S}_2,\mathcal{S}_2)\leq 0$. The enstrophy-maximising irreversible dynamics of any observable $f\in C^\infty(\mathfrak{X}_\mathrm{div}(M))$ is obtained by reversing the sign of the symmetric bracket
\begin{equation}
\frac{d}{dt} f = \{f,\mathcal{H}\} + (f,\mathcal{S}_2),
\end{equation}
and one obtains $\frac{d}{dt} \mathcal{H} = 0$, and $\frac{d}{dt}\mathcal{S}_2 = (\mathcal{S}_2,\mathcal{S}_2) \geq 0$. 

The Hamiltonian is the kinetic energy, and is conserved. 
So if the initial kinetic energy is nonzero, then $\|u\|_\mathfrak{X}\neq0$ for
all time.
The equilibria and long-time behaviour of
the enstrophy-minimising flow are described after
Theorem~\ref{thm:thermodynamic_equilibrium} below. 
The enstrophy-maximising flow, by contrast, has no natural end state: the
enstrophy is unbounded above on energy level sets, so it cannot act as a
Lyapunov functional. Its equations of motion are those derived below with
$\mathrm{Re}$ replaced by $-\mathrm{Re}$, so the dissipative term becomes
$-\mathrm{Re}^{-1}\Delta u^\flat$ plus a nonlocal correction. Equivalently,
this flow is the time reversal of the enstrophy-minimising flow: if $u(t)$
solves the latter, then $-u(-t)$ solves the former. It is therefore of
backward-parabolic type, and we do not expect it to be well posed in the
sense of Hadamard, just as the backward heat equation is not.

For the equations of motion corresponding to the enstrophy-minimising flow, we obtain 
\begin{equation}\label{eq:enstrophyminimising}
    \frac{\partial}{\partial t}u +\left(\mathrm{ad}^*_u u^{\flat}\right)^{\sharp} = \frac{1}{\mathrm{Re}}\Big((\star d\omega)^\sharp +  \frac{\|\omega\|^2_{L^2}}{\|u\|^2_\mathfrak{X}}u\Big), \quad u\in\mathfrak{X}_{\mathrm{div}}(M).
\end{equation}
Since $\omega=\star du^\flat$, the first term on the right hand side is the Laplace-De Rham operator, which follows from the computation $(\star d\star du^\flat)^\sharp = -(d^*du^\flat + dd^*u^\flat)^\sharp = (\Delta u^\flat)^\sharp$, where $dd^*u^\flat =0$ as a result of incompressibility. We emphasise once more that the Laplace-De Rham operator is negative-definite following our conventions. This means that the the enstrophy-minimising irreversible flow corresponds to the Navier-Stokes equations with nonlocal linear friction on the space of divergence-free vector fields over $M$
\begin{equation}\label{eq:NS_energy_conserving}
    \begin{aligned}
        \frac{\partial}{\partial t} u + \left(\mathrm{ad}^*_u u^{\flat}\right)^{\sharp} = \frac{1}{\mathrm{Re}}\Big((\Delta u^\flat)^\sharp +  \frac{\|\omega\|^2_{L^2}}{\|u\|^2_\mathfrak{X}}u\Big), \quad u\in\mathfrak{X}_{\mathrm{div}}(M).
    \end{aligned}
\end{equation}

To pass from $\mathfrak{X}_{\mathrm{div}}(M)$ to the space $\mathfrak{X}(M)$ of all vector fields, we reintroduce the pressure $p\in C^\infty(M)$ as a Lagrange multiplier enforcing the incompressibility constraint. 
Note that both terms on the right-hand side of \eqref{eq:NS_energy_conserving} are already divergence-free, 
and hence the pressure is generated by the advective term alone, exactly as for the Euler equations \eqref{eq:euler}. We obtain the energy-conserving Navier--Stokes equations, with a nonlocal energy-restoring term, in Euler--Arnold form (left) and Euler--Poincar\'e form (right):
\begin{equation}\label{eq:navierstokesgeneral}
    \begin{aligned}
        \frac{\partial}{\partial t} u + \left(\mathrm{ad}^*_u u^{\flat}\right)^{\sharp} &= -(dp)^\sharp +  \frac{1}{\mathrm{Re}}\Big((\Delta u^\flat)^\sharp +  \frac{\|\omega\|^2_{L^2}}{\|u\|^2_\mathfrak{X}}u\Big), & \quad \frac{\partial}{\partial t} u^\flat + \mathcal{L}_u u^\flat &= -dp +  \frac{1}{\mathrm{Re}}\Big(\Delta u^\flat + \frac{\|\omega\|^2_{L^2}}{\|u^\flat\|^2_{L^2}}u^\flat\Big)\\
        \mathrm{div}_\mu u&=0, & \qquad d^* u^\flat&=0.
    \end{aligned}
\end{equation}

For readers familiar with the functional-analytic formulation of the Navier--Stokes equations \cite{temam1979navier}, we note that \eqref{eq:navierstokesgeneral} can be written as an abstract evolution equation on $\mathfrak{X}_{\mathrm{div}}(M)$. Let $A = -\mathbb{P}(\Delta(\cdot)^\flat)^\sharp$ be the Stokes operator defined with the Hodge--de Rham Laplacian, and $B(u,v) = \mathbb{P}\,\nabla_u v$ the convective term, where $\nabla$ denotes the Levi-Civita connection. Since the Hodge--de Rham Laplacian commutes with $d^*$, $\Delta u^\flat$ is divergence-free whenever $u$ is, so $Au = -(\Delta u^\flat)^\sharp$ without projection. Moreover, $\mathcal{L}_u u^\flat = (\nabla_u u)^\flat + \tfrac12 d|u|^2$, so $\mathbb{P}(\mathrm{ad}^*_u u^\flat)^\sharp = B(u,u)$. Finally, $\langle Au,u\rangle_{\mathfrak{X}} = \|du^\flat\|^2_{L^2} = \|\omega\|^2_{L^2}$. Hence \eqref{eq:navierstokesgeneral} is equivalent to
\begin{equation}\label{eq:NS_abstract}
    \frac{du}{dt} + \frac{1}{\mathrm{Re}}\Big(Au - \frac{\langle Au,u\rangle_{\mathfrak{X}}}{\|u\|^2_{\mathfrak{X}}}\,u\Big) + B(u,u) = 0,
    \qquad u \in \mathfrak{X}_{\mathrm{div}}(M).
\end{equation}
The bracketed term is the orthogonal projection of $Au$ onto the tangent space at $u$ of the energy level set $\{\mathcal{H} = \mathcal{H}(u)\}$, in agreement with \eqref{eq:projenstrophy}.

The vorticity formulation is obtained by applying $\star d$ to the Euler-Poincar\'e form (on the right in \eqref{eq:navierstokesgeneral}) of the equations, which yields
\begin{equation}\label{eq:vorticity}
    \frac{\partial}{\partial t}\omega + \mathcal{L}_u\omega = \frac{1}{\mathrm{Re}}\left(\Delta\omega + \frac{\|\omega\|^2_{L^2}}{\|u^\flat\|^2_{L^2}}\omega\right).
\end{equation}
Given the Euler-Poincar\'e form of the velocity equation and the vorticity equation, it is straightforward to establish energy conservation and enstrophy dissipation. By taking the wedge product with $\star u^\flat$ of the Euler-Poincar\'e form of the Navier-Stokes equations and integrating over the domain, we obtain
\begin{equation}
    \frac{1}{2}\frac{\partial}{\partial t}\|u^\flat\|^2_{L^2} = \frac{1}{\mathrm{Re}}\left(-\|\omega\|^2_{L^2} +\frac{\|\omega\|^2_{L^2}}{\|u^\flat\|^2_{L^2}}\|u^\flat\|^2_{L^2}\right) = 0,
\end{equation}
which shows that the energy is conserved. Before we consider the evolution of the enstrophy, we introduce the following lemma.

\begin{lemma}\label{lemma:enstrophy}
Let $(M,\langle\,\cdot\,,\,\cdot\,\rangle)$ be a compact, oriented, Riemannian $2$-manifold without boundary. Let $u$ be a divergence-free vector field with velocity 1-form $u^\flat$ and scalar vorticity $\omega=\star d u^\flat$. Then
\begin{equation}\label{eq:projection_identity}
\Big\|\star d\omega-\frac{\langle \star d\omega,u^\flat\rangle_{L^2}}{\|u^\flat\|_{L^2}^2}\,u^\flat\Big\|_{L^2}^2
= \|d\omega\|_{L^2}^2 - \frac{\|\omega\|_{L^2}^4}{\|u^\flat\|_{L^2}^2},
\end{equation}
where $\langle\alpha,\beta\rangle_{L^2}=\int_M \alpha\wedge\star\beta$ and $\|\cdot\|_{L^2}$ is the associated $L^2$-norm on forms.
\end{lemma}

\begin{proof} In any real inner product space, if we denote the orthogonal projection of $v$ onto $\operatorname{span}(w)$ by
\begin{equation}
P_w(v)\coloneqq\frac{\langle v,w\rangle}{\|w\|^2}\,w,
\end{equation}
then Pythagoras gives
\begin{equation}\label{eq:abstract_pythagoras}
\|v-P_w(v)\|^2 = \|v\|^2 - \|P_w(v)\|^2 = \|v\|^2 - \frac{\langle v,w\rangle^2}{\|w\|^2}.
\end{equation}
Applying this to the space $\Omega^1(M)$ equipped with the $L^2$-inner product, with $v=\star d\omega$ and $w=u^\flat$, yields
\begin{equation}\label{eq:pythagoras_applied}
\Big\|\star d\omega-\frac{\langle \star d\omega,u^\flat\rangle_{L^2}}{\|u^\flat\|_{L^2}^2}\,u^\flat\Big\|_{L^2}^2
= \|\star d\omega\|_{L^2}^2 - \frac{\langle \star d\omega,u^\flat\rangle_{L^2}^2}{\|u^\flat\|_{L^2}^2}.
\end{equation}
Since $\star$ is a pointwise isometry we have $\|\star d \omega\|_{L^2} = \|d \omega\|_{L^2}$. For the second term on the right, we use the identity 
\[\star\alpha\wedge\star\beta=\alpha\wedge\beta,\qquad \alpha,\beta\in\Omega^1(M),\]
which holds since $M$ is a two-dimensional oriented, Riemannian manifold. Integration by parts (with no boundary) together with the identity $du^\flat=\omega\mu$ (from \eqref{eq:starduflat}) gives
\begin{equation}\label{eq:lemmavort}
\langle \star d\omega,u^\flat\rangle_{L^2}
= \int_M \star d\omega\wedge\star u^\flat
= \int_M d\omega\wedge u^\flat
= -\int_M \omega\,du^\flat
= -\int_M \omega^2\mu
= -\|\omega\|_{L^2}^2.
\end{equation}
Substituting \eqref{eq:lemmavort} into \eqref{eq:pythagoras_applied} yields the result.
\end{proof}
To obtain the rate of dissipation of the enstrophy, we take the wedge product of the vorticity equation \eqref{eq:vorticity} with $\star\omega$ and integrate over the domain. By means of Lemma \ref{lemma:enstrophy}, we obtain
\begin{equation}\label{eq:enstrophy_dissipation}
    \frac{1}{2}\frac{\partial}{\partial t}\|\omega \|^2_{L^2} = \frac{1}{\mathrm{Re}}\left(-\|d\omega\|^2_{L^2} + \frac{\|\omega\|^4_{L^2}}{\|u^\flat\|^2_{L^2}}\right) = -\frac{1}{\mathrm{Re}}\Big\|\star d\omega-\frac{\langle \star d\omega,u^\flat\rangle_{L^2}}{\|u^\flat\|_{L^2}^2}\,u^\flat\Big\|_{L^2}^2\leq 0.
\end{equation}
The right hand side of the evolution equation for the enstrophy implies the inequality
\begin{equation}\label{eq:the_CS_inequality}
    \|\omega\|^4_{L^2} \leq \|u^\flat\|^2_{L^2}\|d\omega\|^2_{L^2},
\end{equation}
which is a direct consequence of the Cauchy-Schwarz inequality applied to $\langle \star d\omega,u^\flat\rangle_{L^2}$. The next theorem shows that the condition for equality is equivalent to the vanishing of the dissipative contribution in Equation \eqref{eq:navierstokesgeneral} and also yields stationary solutions of the Euler equations.
\begin{theorem}\label{thm:thermodynamic_equilibrium}
Suppose $u$ is a smooth, non-zero, divergence-free vector field on $M$. Then the following statements are equivalent:
\begin{enumerate}
  \item Equality holds in \eqref{eq:the_CS_inequality}.
  \item The stream function $\psi$ is an eigenfunction of the Laplace-Beltrami operator:
  \[
    \Delta\psi=-c(u)\psi.
  \]
  \item The velocity $1$-form $u^\flat$ is an eigenform of the Hodge-de Rham Laplacian:
  \[
    \Delta u^\flat=-c(u)u^\flat.
  \]
  \item The terms arising from the symmetric bracket contribution in
  \eqref{eq:navierstokesgeneral} vanishes, i.e. \[\frac{1}{\mathrm{Re}}\Big(\Delta u^\flat + \frac{\|\omega\|^2_{L^2}}{\|u^\flat\|^2_{L^2}}u^\flat\Big) = 0.\]
\end{enumerate}
Moreover, whenever these conditions hold, $u$ is a stationary solution of both the Euler equations and
the metriplectic equations.
\end{theorem}

\begin{proof}
Since $u$ is nonzero, divergence-free, and $M$ is compact and simply-connected, one has $\omega\neq0$. In case of the torus, restricting to flows with zero mean velocity guarantees this. Indeed, if $\omega=0$, then
$du^\flat=0$, while incompressibility gives $d^\ast u^\flat=0$.
Thus $u^\flat$ would be harmonic and hence zero, since
$M$ is simply-connected. Consequently, $c(u)>0$.

\medskip

\noindent
$(1)\implies(2)$.
By Lemma \ref{lemma:enstrophy}, equality in the Cauchy--Schwarz inequality
holds precisely when
\[
  \star d\omega
  =
  \frac{\langle\star d\omega,u^\flat\rangle_{L^2}}
       {\|u^\flat\|_{L^2}^2}\,u^\flat.
\]
Since
\[
  \langle\star d\omega,u^\flat\rangle_{L^2}
  =-\|\omega\|_{L^2}^2,
\]
this condition is equivalent to
\begin{equation}\label{eq:equality-condition}
  \star d\omega=-c(u)u^\flat.
\end{equation}
Using $u^\flat=\star d\psi$ and the fact that $\star$ is an isomorphism
on $\Omega^1(M)$, we obtain
\[
  d\omega=-c(u)d\psi.
\]
Hence
\[
  \omega=-c(u)\psi+C
\]
for some spatial constant $C$. We may normalise $\psi$ to have zero
mean, since it is determined only up to an additive constant. Moreover,
by Stokes' theorem,
\[
  \int_M\omega\,\mu=\int_Mdu^\flat=0.
\]
Integrating the preceding relation over $M$ therefore gives $C=0$.
Since $\omega=\Delta\psi$, it follows that
\[
  \Delta\psi=-c(u)\psi.
\]

\medskip
\noindent
$(2)\implies(3)$.
Since the Hodge-de Rham Laplacian commutes with both $d$ and $\star$, we obtain
\[
  \Delta u^\flat
  =\Delta(\star d\psi)
  =\star d(\Delta\psi).
\]
Then $\Delta\psi=-c(u)\psi$ implies that
\[
  \Delta u^\flat
  =-c(u)\star d\psi
  =-c(u)u^\flat.
\]

\medskip
\noindent
$(3)\implies(4)$.
The symmetric bracket contribution in
\eqref{eq:navierstokesgeneral}, written in terms of the velocity
$1$-form, is
\[
  \frac{1}{\mathrm{Re}}
  \left(\Delta u^\flat+c(u)u^\flat\right).
\]
If $\Delta u^\flat=-c(u)u^\flat$, this contribution vanishes.

\medskip
\noindent
$(4)\implies(1)$.
If the symemtric bracket contribution vanishes, then
\[
  \Delta u^\flat+c(u)u^\flat=0.
\]
Since $\Delta u^\flat=\star d\omega$, this is equivalent to
\[
  \star d\omega=-c(u)u^\flat.
\]
Thus $\star d\omega$ is a scalar multiple of $u^\flat$, and equality
holds in the Cauchy-Schwarz inequality. This completes the proof of
the equivalence.

\medskip
Finally, suppose $(2)$ holds. Then
\[
  \omega=\Delta\psi=-c(u)\psi,
\]
and hence
\[
  \{\psi,\omega\}_{\mathrm{can}}
  =-c(u)\{\psi,\psi\}_{\mathrm{can}}
  =0.
\]
Thus $u$ is a stationary solution of the Euler equations. By the equivalence
proved above, statement $(2)$ also implies statement $(4)$, so the
symmetric bracket contribution in \eqref{eq:navierstokesgeneral} vanishes.
Therefore, $u$ is also a steady solution of the metriplectic equations
\eqref{eq:navierstokesgeneral}.
\end{proof}

 The condition for equality is the two-dimensional analogue of a Beltrami (vorticity aligned) flow. The inequality \eqref{eq:the_CS_inequality} also provides a bound on the coefficient $c(u)$, namely
\begin{equation}
    c(u)\leq \frac{\|d\omega\|_{L^2}}{\|u^\flat\|_{L^2}},
\end{equation}
which shows that, for every nonzero $u$, the coefficient $c(u)$ is finite. This inequality controls how large the enstrophy-to-energy ratio can be in terms of $\|d\omega\|_{L^2}$. 

\begin{proposition}\label{prop:poincare}
Let $0<\lambda_1\le\lambda_2\le\cdots$ be the nonzero eigenvalues of
$-\Delta$ on $M$. For every nonzero $u\in\mathfrak{X}_{\mathrm{div}}(M)$,
\begin{equation}\label{eq:poincare_enstrophy}
    \|\omega\|^2_{L^2}\;\ge\;\lambda_1\,\|u\|^2_\mathfrak{X},
    \qquad\text{equivalently}\qquad c(u)\ge\lambda_1,
\end{equation}
with equality if and only if $\psi$ lies in the first eigenspace.
\end{proposition}

\begin{proof}
Since $M$ is simply connected, $u^\flat=\star d\psi$, and we may take $\psi$
to have zero mean. Expanding $\psi=\sum_k a_k\phi_k$ in $L^2$-orthonormal
eigenfunctions of $-\Delta$ gives
$\|u\|^2_\mathfrak{X}=\|d\psi\|^2_{L^2}=\sum_k\lambda_k a_k^2$, via integration by parts. Moreover, since
$\|\omega\|^2_{L^2}=\|\Delta\psi\|^2_{L^2}=\sum_k\lambda_k^2a_k^2$, we have \[\|\omega\|^2_{L^2} - \lambda_1\|u\|^2_\mathfrak{X} = \sum_k \lambda_k(\lambda_k - \lambda_1)a_k^2\geq 0\] with equality if and only if $a_k = 0$ for all $k$ with $\lambda_k>\lambda_1$, or equivalently, $\psi$ lies in the first eigenspace.
\end{proof}

Together with \eqref{eq:the_CS_inequality}, this gives
$\lambda_1\le c(u)\le\|d\omega\|_{L^2}/\|u\|_\mathfrak{X}$, and at the
equilibria of Theorem~\ref{thm:thermodynamic_equilibrium} one has
$c(u)=\lambda_k$ for some $k$. Since the energy is conserved, along the
enstrophy-minimising flow
\[
    \|\omega(t)\|^2_{L^2}\;\ge\;\lambda_1\,\|u(0)\|_\mathfrak{X}^2\;>\;0
    \qquad\text{for all } t,
\]
so the enstrophy is bounded away from zero, and the flow cannot approach an
irrotational state. The enstrophy decreases monotonically and is bounded
below, so it converges. Its minimisers on an energy level set are the
gravest (largest-scale,
lowest-eigenvalue) eigenmodes, while the higher eigenmodes are saddle points of
$\mathcal{S}_2$ restricted to that level set. The gravest mode is therefore
the expected generic end state, in the spirit of the minimum-enstrophy states
of \cite{bretherton1976two}. We do not prove convergence here: this would
require a LaSalle-type argument, together with compactness of trajectories in
a suitable topology.

\begin{remark}\label{rem:BDM}
On the flat torus $\mathbb{T}^2$, restricted to mean-zero fields and to the energy level $\|u\|_{\mathfrak{X}}=1$, equation~\eqref{eq:NS_abstract} with $\nu=1/\mathrm{Re}$ becomes
\begin{equation}\label{eq:BDM}
    \frac{du}{dt} + \nu A u - \nu\|\nabla u\|_{L^2}^2\,u + B(u,u) = 0, \qquad \|u\|_{L^2}=1.
\end{equation}
This is the \emph{constrained Navier--Stokes equation} of \cite{brzezniak2018constrained}; here we used $\langle Au,u\rangle_{\mathfrak{X}} = \|\nabla u\|^2_{L^2}$ on $\mathbb{T}^2$. The authors obtain~\eqref{eq:BDM} by projecting the Navier--Stokes equations onto the tangent space of the sphere $\mathcal{M}=\{\|u\|_{L^2}=1\}$, writing it as $\dot u = -\nabla_{\mathcal{M}} E(u) - B(u,u)$ with $E(u)=\tfrac12\|\nabla u\|_{L^2}^2$. In our terms, $\mathcal{M}$ is a level set of $\mathcal{H}$, $E=\mathrm{Re}\,\mathcal{S}_2$ and $\nabla_{\mathcal{M}} E = \mathrm{Re}\,P^\perp_{\mathcal{H}}\,\delta\mathcal{S}_2/\delta u$. The orthogonality $\langle B(u,u),Au\rangle_{\mathfrak{X}}=0$ underlying their global existence proof expresses the conservation of $\mathcal{S}_2$ by the Euler flow $\dot u = -B(u,u)$: since $\mathcal{S}_2$ is a Casimir and $\delta\mathcal{S}_2/\delta u = \mathrm{Re}^{-1}Au$,
\begin{equation*}
    0 = \{\mathcal{S}_2,\mathcal{H}\} = \Big\langle \frac{\delta\mathcal{S}_2}{\delta u},\,-B(u,u)\Big\rangle_{\mathfrak{X}} = -\frac{1}{\mathrm{Re}}\,\langle Au,\,B(u,u)\rangle_{\mathfrak{X}},
\end{equation*}
so that in~\eqref{eq:BDM} the advective term does not contribute to the evolution of $\mathcal{S}_2$. The resulting monotone decay of $\|\nabla u\|_{L^2}$, which they use to extend local solutions globally, is~\eqref{eq:enstrophy_dissipation}.

Two differences are worth noting. First, \eqref{eq:BDM} conserves energy only on $\mathcal{M}$, while
for~\eqref{eq:NS_abstract}, 
every energy level is invariant. Second, on a curved surface $\langle Au,u\rangle_{\mathfrak{X}} = \|\omega\|^2_{L^2}$ differs from $\|\nabla u\|^2_{L^2}$ by a curvature term.

In \cite{brzezniak2018constrained}, the existence of global solutions of~\eqref{eq:BDM} is proved on $\mathbb{T}^2$, together with convergence to the Euler equations in the vanishing-viscosity limit. The authors note that their global argument fails on general bounded domains, where the orthogonality $\langle B(u,u),Au\rangle_{\mathfrak{X}}=0$ is lost. Since this orthogonality follows from the Casimir property, it holds on any closed surface, which suggests that their results extend to this setting.
In three dimensions, by contrast, vortex stretching destroys the
conservation of enstrophy, so the orthogonality fails and the argument
does not extend; see Remark~\ref{rem:3D-NS}.
\end{remark}

\subsubsection{General vorticity moments}
In the previous section, we showed that the irreversible dynamics associated with the enstrophy produces an energy-preserving version of the Navier-Stokes equations in two dimensions. In this section, we show that using general vorticity moments to determine the symmetric bracket leads to nonlinear diffusion balanced by nonlocal terms. Let $\mathcal{S}_p(\omega) = \frac{1}{p}\int_M|\omega|^p \mu$ with $p\geq 2$ (one may drop the absolute value signs for even moments $p=2n$ for $n\in\mathbb{N})$. It follows that
\begin{equation}
    \frac{\delta\mathcal{S}_p}{\delta u} = -(\star d(|\omega|^{p-2}\omega))^\sharp=-(p-1)|\omega|^{p-2}(\star d\omega)^\sharp,
\end{equation}
and the projection of the variational derivative onto the orthogonal direction of the variational derivative of the Hamiltonian yields
\begin{equation}
    P_\mathcal{H}^\perp \frac{\delta \mathcal{S}_p}{\delta u} = -(p-1)|\omega|^{p-2}(\star d\omega)^\sharp - \frac{\int_M|\omega|^p \mu}{\|u\|^2_\mathfrak{X}}u.
\end{equation}
Taking instead the $L^p$ norm of $\omega$ as the Casimir function, one obtains the same dynamics up to a reparametrisation of time, since $\delta\|\omega\|_{L^p} = \|\omega\|^{1-p}_{L^p}\delta\mathcal{S}_p$ for $\omega\neq 0$. The Casimir-minimising flow is energy-preserving and the equation of motion is given in Euler-Poincar\'e form by
\begin{equation}
    \begin{aligned}
    \frac{\partial}{\partial t}u^\flat + \mathcal{L}_u u^\flat &= -d\pi +(p-1) |\omega|^{p-2}\Delta u^\flat + \frac{\int_M|\omega|^p\mu}{\|u^\flat\|_{L^2}^2}u^\flat,\\
    d^* u^\flat &= 0,
    \end{aligned}
\end{equation}
where we have denoted the pressure by $\pi$ to avoid confusion with the exponent $p$. The case for $p=2$ corresponds to the enstrophy and is the only choice producing linear diffusion in the moment hierarchy. For $p>2$, the diffusion is nonlinear and degenerate, it vanishes where $\omega=0$. The diffusion term acts locally and for $p>2$, regions of high vorticity experience stronger dissipation. The accompanying nonlocal term acts globally and ensures that the irreversible deformation of the flow stays on kinetic energy surfaces.

\section{Irreversible fluid dynamics in 3D}\label{sec:threedimensions}

The irreversible dynamics of a purely incompressible fluid flow in three dimensions cannot be treated in the same way as in the two-dimensional case discussed in the previous section. The essential reason is that a three-dimensional incompressible fluid without additional advected quantities possesses only a single regular Casimir invariant, the helicity \cite{enciso2016helicity}, rather than an uncountable family.

In any dimension, the Euler flow advects the vorticity 2-form, 
i.e. $(\partial_t+L_u)\,du^\flat=0$, and preserves the volume form $\mu$. In two dimensions $du^\flat$ is of top degree, so $du^\flat=\omega\,\mu$ for a scalar vorticity $\omega$, which is therefore itself advected;
every functional $\int_M\Phi(\omega)\,\mu$ is then conserved. In three dimensions, this mechanism fails: $du^\flat$ is not of top degree, so extracting a scalar from it requires additional
structure. The metric provides one, through the pointwise norm $|du^\flat|$,
but this norm is not advected, because the flow preserves volumes
but not lengths. In physical terms: vortex lines are carried along with
the fluid, but as fluid elements are deformed, the vortex lines through
them are stretched and tilted, changing the magnitude and direction 
of the vorticity
vector $\omega:=(\star du^\flat)^\sharp$. Hence, for example, the enstrophy $\Omega \coloneqq \tfrac12\int_M|\omega|^2\mu$ is not
conserved. This \textit{vortex stretching} has no two-dimensional
counterpart, and it removes the infinite family of Casimirs found in two dimensions. The only
top-degree form that can be built from 
$u^\flat$ alone, using only the exterior derivative and
wedge product, is the
helicity 3-form $u^\flat\wedge du^\flat$, whose integral is the single
remaining Casimir; see Section~\ref{ssect:helicity}.


To obtain a richer set of invariants in three dimensions, one may introduce an additional advected scalar field $T$, such as buoyancy or temperature, and combine it with the vorticity 2-form to produce a scalar ``vorticity-like'' quantity: the Ertel potential vorticity $q$ defined by
$q\,D\mu = du^\flat\wedge dT$,
where $D$ is the mass density; see Section~\ref{ssect:PV}.
The 3-form $du^\flat\wedge dT$ is advected
(Proposition~\ref{prop:PVadvected}), and so is the mass form $D\mu$;
hence $q$ is advected, exactly as the scalar vorticity is in two
dimensions, and so every functional $\int_M\Phi(q)\,D\mu$ is a Casimir
(Proposition~\ref{prop:ErtelCasimir}).
Potential vorticity thus plays in three dimensions the structural role that the scalar vorticity plays in two,
providing an uncountable family of Casimir invariants and enabling the construction of irreversible dynamics.

\begin{remark}[Energy-conserving Navier--Stokes in three dimensions]
\label{rem:3D-NS}
It is in fact possible to proceed as in the two-dimensional case,
using the enstrophy $\Omega:=\frac{1}{2\,\mathrm{Re}}\int_M|\omega|^2\mu=\frac{1}{2\,\mathrm{Re}}\|du^\flat\|_{L^2}^2$.
Since $\Omega$ is not conserved by the Euler
flow, 
it is not admissible as a
generalised entropy in the sense of Section~\ref{sec:generaltheory}.
Nevertheless we proceed by defining the dynamics
\[
  \frac{d}{dt}f=\{f,H\}-(f,\Omega)_H,
\]
where $(\cdot, \cdot)_H$ is the symmetric bracket defined in \eqref{eq:symK}.
This equation is almost the two-generator form in \eqref{eq:twogen} 
except that we have omitted the term $-\{f, \Omega\}$, which is in general nonzero.
The dynamics still conserves $H$, since this requires only the
degeneracy of the symmetric bracket. 
On divergence-free fields,
$\delta \Omega/\delta u=\frac{1}{\mathrm{Re}}(d^*du^\flat)^\sharp
=-\frac{1}{\mathrm{Re}}(\Delta u^\flat)^\sharp$, and so exactly as in
Section~\ref{sssect:enstrophy} one obtains
\[
  \frac{\partial}{\partial t}u^\flat+L_uu^\flat
  =-dp+\frac{1}{\mathrm{Re}}\Big(\Delta u^\flat
   +\frac{\|du^\flat\|_{L^2}^2}{\|u^\flat\|_{L^2}^2}\,u^\flat\Big),
  \qquad d^*u^\flat=0,
\]
the three-dimensional analogue of~\eqref{eq:navierstokesgeneral}: the
Navier--Stokes equations (with the Hodge Laplacian) together with a
nonlocal term restoring energy balance. What is lost is monotonicity of $\Omega$,
since the rate of change of $\Omega$ along the Euler
flow has no definite sign (and is independent of $\mathrm{Re}$). 
The system is therefore not metriplectic, and
$\Omega$ is not a Lyapunov functional. 
This is why
we work instead with the potential-vorticity Casimirs introduced above,
which retain a monotone entropy.
\end{remark}

Let $(M,\langle\,\cdot\,,\,\cdot\,\rangle)$ be a three-dimensional compact, orientable Riemannian manifold without boundary with Riemannian volume form $\mu$. Instead of restricting attention to the group of volume-preserving diffeomorphisms, we consider the full diffeomorphism group $\mathrm{Diff}(M)$. Following the Euler-Poincar\'e variational principle of \cite{holm1998euler}, we impose incompressibility using Lagrange multipliers rather than constraining the configuration space a priori. In this formulation, the mass density becomes an advected quantity and is therefore allowed to evolve. We also include an additional scalar field. 

Both the density and the scalar field are advected by the flow, and it follows that the natural configuration space is the semidirect product group
\begin{equation}
    \mathrm{Diff}(M)\ltimes(C^\infty(M)\times \mathrm{Dens}_\mu(M))
\end{equation}
where each component transforms under pullback by diffeomorphisms. This semidirect product structure encodes the advection of the scalar field and the density field, as well as the forces associated with these advected quantities. The Lie algebra associated with this semidirect product is 
\begin{equation}
    \mathfrak{S} = \mathfrak{X}(M)\ltimes (C^\infty(M)\times \mathrm{Dens}_\mu(M))
\end{equation}
with the infinitesimal action given by the Lie derivative on each component
\begin{equation}
    u\cdot(\nu\mu,\xi) = (\mathcal{L}_u(\nu\mu),\mathcal{L}_u\xi).
\end{equation}
(Note that the the
physical quantities will appear as the dual variables below.)
The Lie bracket on the semidirect product Lie algebra then has the form
\begin{equation}\label{eg:semiLieBracket}
    [(u,\nu\mu,\xi),(v,\nu'\mu,\xi')] = ([u,v]_\mathfrak{X},\ \mathcal{L}_u(\nu'\mu)-\mathcal{L}_v(\nu\mu),\ \mathcal{L}_u\xi'-\mathcal{L}_v\xi).
\end{equation}
The fluid now can transport both advected fields independently, so the natural functional describing the dynamics is more than just the kinetic energy associated to the weak Riemannian metric on $\mathrm{Diff}(M)$. Indeed, the typical Lagrangian functional contains the kinetic energy and potential energy terms associated with the advected quantities. The dual space of $\mathfrak{S}$ is the space $\mathfrak{S}^*\supset\mathfrak{X}^*(M)\ltimes(C^\infty(M)\times\mathrm{Dens}_\mu(M))$ with respect to the pairing
\begin{equation}\label{eq:pairing3d}
    \Big\langle(m\otimes \mu,T,D\mu),(u,\nu\mu,\xi)\Big\rangle_{\mathfrak{S}^*\times\mathfrak{S}}
    = \int_M \iota_u (m \otimes \mu) + \int_M T\nu\,\mu + \int_M D\xi\,\mu,
\end{equation}
where $m \otimes \mu \in\mathfrak{X}^*(M)$ is the covector-valued density with coefficient $1$-form $m$,
conjugate to the velocity field $u$;
the advected scalar field $T\in C^\infty(M)$ is
conjugate to the density component $\nu\mu$; and the mass form
$D\mu\in\mathrm{Dens}_\mu(M)$, with scalar coefficient $D$, is conjugate to the scalar component $\xi$.
The mass density coefficient $D$ 
is denoted $\rho$ in the classical fluid dynamics literature; 
we follow the naming convention of \cite{holm1998euler}.
This pairing is linear in both entries and provides the natural setting for defining Hamiltonian functionals and the corresponding Lie-Poisson bracket on the dual of the semidirect product Lie algebra. 

\begin{remark}[Densities versus their coefficients]\label{rem:coefficient-convention}
    Every element of $\mathrm{Dens}_\mu(M)$ 
    can be uniquely expressed as 
    $f\,\mu$ for some coefficient $f \in C^\infty(M)$. In particular, $D\in C^\infty(M)$ is the mass
    density coefficient and $D\mu\in\mathrm{Dens}_\mu(M)$ is the mass form. 
    Similarly, $m \coloneqq D\,u^\flat\in\Omega^1(M)$ is the momentum density coefficient
    $1$-form, while
    $m\otimes\mu\in\mathfrak{X}^*(M)\cong\Omega^1(M)\otimes\mathrm{Dens}_\mu(M)$
    is the covector-valued density dual to $u$.
    The distinction between densities and their coefficients is not merely notational, since the two objects
    transport differently: for any $\beta\in\Omega^k(M)$,
    \begin{align}\label{eq:coefficient-retyping}
    \mathcal{L}_u(\beta\otimes\mu)
    = \bigl(\mathcal{L}_u\beta + (\operatorname{div}_\mu u)\,\beta\bigr)\otimes\mu, 
    \end{align}
In the classical fluid dynamics literature the coefficients $D$ and $m$
    are written $\rho$ and $\rho\mathbf{u}$ respectively; we follow the
    naming convention of \cite{holm1998euler}.
\end{remark}

\begin{remark}\label{rem:slots}
    Physically, in line with Tonti's thermodynamic classification \cite{tonti1976reason, tonti2013mathematical}, scalar advected quantities model \emph{intensive} quantities
    describing the local state of a fluid parcel independently of its size, such as temperature, buoyancy, or specific entropy.
    
    Instead of a scalar advected quantity $T$, the framework equally supports an advected density $\widetilde{T}\mu\in\mathrm{Dens}_\mu(M)$. Densities are the natural objects that integrate over material regions to yield \emph{extensive} totals, e.g.\ $\int_\Omega D\mu$ is the mass of the region $\Omega$. If $\widetilde{T} = DT$, i.e. a mass-weighted counterpart of $T$, then $\widetilde{T}\mu$ is the density whose integral $\int_\Omega \widetilde{T}\mu$ gives the total of $T$ over $\Omega$, weighted by mass -- e.g.\ for $T=s$ specific entropy, $\widetilde{T}=Ds$ is the entropy density and $\int_\Omega Ds\,\mu$ the total entropy in $\Omega$. (Note that the mass-weighted  counterparts of temperature and buoyancy do not have standard physical interpretations.) We fix the scalar convention throughout, i.e.\ take $T$ to be a scalar advected quantity, as required by the Ertel construction.
\end{remark}

The Lie--Poisson bracket on $C^\infty(\mathfrak{S}^*)$ is directly obtained from the Jacobi--Lie bracket on $\mathfrak{S}$ (with a minus sign, following the right-invariance of the Lie--Poisson bracket)
\eqref{eg:semiLieBracket} by substituting the functional derivative
$\big(\delta\mathcal{F}/\delta m,\ (\delta\mathcal{F}/\delta T)\,\mu,\
\delta\mathcal{F}/\delta D\big)\in\mathfrak{S}$ — whose density component
has coefficient function $\delta\mathcal{F}/\delta T$ and whose scalar
component is $\delta\mathcal{F}/\delta D$, consistently with the pairing
\eqref{eq:pairing3d} — and pairing against $(m,T,D\mu)$. This gives
\begin{equation}\label{eq:lpbracket3d_densities}
    \begin{aligned}
    \{\mathcal{F},\mathcal{G}\}_\mathfrak{S} &= -\int_M{\left[\frac{\delta\mathcal{F}}{\delta m},\frac{\delta \mathcal{G}}{\delta m}\right]_\mathfrak{X}}\contract m 
    + \int_M T\left(\mathcal{L}_{\delta\mathcal{F}/\delta m}\Big(\frac{\delta\mathcal{G}}{\delta T}\,\mu\Big)
    -\mathcal{L}_{\delta \mathcal{G}/\delta m}\Big(\frac{\delta \mathcal{F}}{\delta T}\,\mu\Big)\right) \\
    &\quad + \int_M D\left(\mathcal{L}_{\delta\mathcal{F}/\delta m}\Big(\frac{\delta\mathcal{G}}{\delta D}\Big)-\mathcal{L}_{\delta \mathcal{G}/\delta m}\Big(\frac{\delta \mathcal{F}}{\delta D}\Big)\right) \mu,
    \end{aligned}
\end{equation}
where the hook notation $u\contract m$ is alternative notation for the interior product $\iota_u m$. 
Note that in the $T$-term (i.e. the second integral)
the Lie derivative acts on densities, so its output is already a top
form and no additional factor of $\mu$ appears. 
To express the $T$-term in terms of coefficient functions instead of densities, we apply  
the identity
$\mathcal{L}_u(f\mu) = \mathrm{div}_\mu(fu)\,\mu$ for
$f\in C^\infty(M)$, $u\in\mathfrak{X}(M)$, together with Stokes'
theorem on the closed manifold $M$:
\begin{equation}
    \int_M T\,\mathcal{L}_u(f\mu)
    = \int_M T\,\mathrm{div}_\mu(fu)\,\mu
    = -\int_M f\,(\mathcal{L}_u T)\,\mu,
\end{equation}
applied with $f = \delta\mathcal{G}/\delta T$,
$u = \delta\mathcal{F}/\delta m$ and with $\mathcal{F},\mathcal{G}$
interchanged. 
This yields
\begin{equation}\label{eq:lpbracket3d}
    \begin{aligned}
    \{\mathcal{F},\mathcal{G}\}_\mathfrak{S} &= -\int_M {\left[\frac{\delta\mathcal{F}}{\delta m},\frac{\delta \mathcal{G}}{\delta m}\right]_\mathfrak{X}}\contract m
    - \int_M \left(\frac{\delta\mathcal{G}}{\delta T}\,\mathcal{L}_{\delta\mathcal{F}/\delta m}T-\frac{\delta\mathcal{F}}{\delta T}\,\mathcal{L}_{\delta \mathcal{G}/\delta m}T\right)\mu \\
    &\quad + \int_M D\left(\mathcal{L}_{\delta\mathcal{F}/\delta m}\Big(\frac{\delta\mathcal{G}}{\delta D}\Big)-\mathcal{L}_{\delta \mathcal{G}/\delta m}\Big(\frac{\delta \mathcal{F}}{\delta D}\Big)\right) \mu.
    \end{aligned}
\end{equation}

A typical Hamiltonian for compressible fluids in three dimensions with an advected scalar field is
\begin{equation}\label{eq:ham3d}
    \mathcal{H}(m,D,T) = \int_M\Big(\frac{1}{2D}\langle m,m\rangle + U(D,T)\Big)\mu,
\end{equation}
where $U(D,T)$ describes the internal energy per unit volume of the fluid. The internal energy dictates whether a compressible fluid is well-approximated by an incompressible fluid. Indeed, the associated pressure is given by the thermodynamic relation
\begin{equation}
p(D,T) = D\frac{\partial U}{\partial D}-U,
\end{equation}
so the fluid’s resistance to compression is governed by the derivative $\partial p/\partial D$, the square of the speed of sound. When this derivative is large, even small density variations produce large pressure forces that rapidly suppress compressibility effects. From this viewpoint, incompressibility can be interpreted as a singular thermodynamic limit in which the pressure reacts infinitely fast to density changes. That is,
\begin{equation}
\frac{\partial p}{\partial D}\to\infty,
\end{equation}
or, in physical terms, the speed of sound becomes infinite. In this limit the density becomes energetically constrained, density variations are forbidden, and the pressure ceases to derive from $U$. Instead, for uniform density, it appears as a Lagrange multiplier enforcing the constraint $d^*u^\flat = 0$ in the incompressible equations.

In geophysical applications, fluids cover a huge range of spatial and temporal scales. While effects of compressibility at small scales cannot be ignored for air, air at large spatial scales is typically modelled as incompressible. The internal energy should then not depend on $D$, i.e. $U(D,T)=U(T)$. The internal energy is often taken quadratic in the passive scalar $T$. If the potential energy depends explicitly on position, one finds internal waves in the dynamics. Furthermore, at planetary scales the Coriolis effect is essential. Introducing a rotating frame can be understood geometrically as a (topologically trivial) central extension or cocycle of the algebra of vector fields. This modification does not change the Hamiltonian functional on $\mathfrak S^*$, but it does change the identification between the momentum $m$ and the velocity \(u\) (equivalently, it produces a steady background momentum or momentum shift). In practice this shift is responsible for the Coriolis terms that appear in the equations of motion. 

Given the Hamiltonian \eqref{eq:ham3d} with internal energy and the Lie-Poisson bracket \eqref{eq:lpbracket3d}, the compressible equations of motion are given by
\begin{equation}\label{eq:compressible}
\begin{aligned}
\left(\frac{\partial}{\partial t}  +\mathcal{L}_{u}\right) (m \otimes \mu) 
&= \left(- D\,d\Big(\frac{\partial U}{\partial D}-\frac12\langle u^\flat,u^\flat\rangle\Big)+\frac{1}{D}\frac{\partial U}{\partial T}\,dT\right) \otimes \mu, &&\qquad \textrm{(momentum)}\\ 
\left(\frac{\partial}{\partial t}  +\mathcal{L}_{u}\right)T &=0,&&\qquad \textrm{(advection)}\\
\left(\frac{\partial}{\partial t}  +\mathcal{L}_{u}\right)(D\mu) &=0, &&\qquad \textrm{(continuity)}
\end{aligned}
\end{equation}
where the partial derivatives of the internal energy function $U$ with respect to $T$ and $D$ provide the force 
exerted by the advected scalar
field $T$ and the pressure, respectively. 

Since $m \otimes \mu = (Du^\flat) \otimes \mu =  u^\flat \otimes D\mu$,
and $\partial_t+\mathcal{L}_u$ is a derivation, we have
\[
    \left(\frac{\partial}{\partial t}  +\mathcal{L}_{u}\right) \left(m \otimes \mu\right)
    = \Big[\left(\frac{\partial}{\partial t}  +\mathcal{L}_{u}\right)u^\flat\Big]\otimes D\mu
      \;+\; u^\flat\otimes\Big[\Big(\frac{\partial}{\partial t}+\mathcal{L}_u\Big)(D\mu)\Big],
\]
and the second term on the right vanishes by the continuity equation.
So the momentum equation is equivalent to
\[
    \Big[\Big(\frac{\partial}{\partial t}+\mathcal{L}_u\Big)u^\flat\Big]\otimes D\mu
    = \left(-D\,d\Big(\frac{\partial U}{\partial D}-\frac12\langle u^\flat,u^\flat\rangle\Big)
         +\Big(\frac{\partial U}{\partial T}\Big) \, dT\right)  \otimes\mu.
\]
Dividing by $D$ (which is always positive) gives the equivalent velocity form of the momentum equation,
\begin{equation}\label{eq:compressible-velocity}
    \Big(\frac{\partial}{\partial t}+\mathcal{L}_u\Big)u^\flat
    = -\,d\Big(\frac{\partial U}{\partial D}-\frac12\langle u^\flat,u^\flat\rangle\Big)
      +\frac{1}{D}\,\Big(\frac{\partial U}{\partial T}\Big) \, dT.
\end{equation}

In case Coriolis forces do not come into play, we formulate the incompressible equations in terms of the velocity $1$-form $u^\flat$, so that the incompressible equations are given by
\begin{equation}\label{eq:incompressible}
    \begin{aligned}
        \frac{\partial}{\partial t} u^\flat + \mathcal{L}_u u^\flat &= -d\Big(p-\frac{1}{2}\langle u^\flat,u^\flat\rangle\Big) + \frac{1}{D_0}\frac{\partial U}{\partial T}\,dT,\\
        \frac{\partial}{\partial t} T +\mathcal{L}_{u}T &=0,\\
        d^*u^\flat &= 0,
    \end{aligned}
\end{equation}
where $D_0$ is the constant mass-density coefficient and
$p$ is the Lagrange multiplier enforcing the incompressibility
constraint, playing the role that $\partial U/\partial D$ plays in
\eqref{eq:compressible}.
If Coriolis forces do play a role, then \eqref{eq:incompressible} would involve extra terms. Since these terms would fundamentally change the treatment of potential vorticity in the next section, we omit them.

\subsection{Potential vorticity}\label{ssect:PV}
The kernel of the Lie--Poisson bracket \eqref{eq:lpbracket3d} is classically expressed in terms of the Ertel potential vorticity $q$, 
defined geometrically as follows.
Define the
\emph{mass-weighted Hodge star} on functions,
\begin{equation}\label{eq:starD}
    \star_D : \Omega^0(M)\to\Omega^3(M),
    \qquad \star_D f \coloneqq f\,D\mu = D\star f .
\end{equation}
Note that since $D>0$, the map $\star_D$ is an isomorphism, with inverse
\begin{equation}\label{eq:starDinv}
    \star_{1/D} : \Omega^3(M)\to\Omega^0(M),
    \qquad 
    \star_{1/D} (g \mu) \coloneqq \frac{g}{D} = \frac{1}{D} \star (g \mu).
\end{equation}
Then define the 
\emph{Ertel potential vorticity} $q\in C^\infty(M)$ by
\begin{align}\label{eq:potentialvorticity}
q &= \star_{1/D} \left(d\big(u^\flat\big)\wedge dT\right), \quad \textrm{i.e.} \\
\star_D q &= d\big(u^\flat\big)\wedge dT = d\big(\frac{m}{D}\big)\wedge dT.
\end{align}
Note that $du^\flat$ is, by definition, the vorticity 2-form. 
\begin{proposition}\label{prop:PVadvected}
For the reversible system \eqref{eq:compressible},
    the potential vorticity $q$ is advected, i.e.
    \begin{equation}
        \frac{\partial}{\partial t}q + \mathcal{L}_u q = 0.
    \end{equation}
\end{proposition}

\begin{proof}
We apply $d$ to both sides of \eqref{eq:compressible-velocity}.
Since $d$ commutes with $\partial_t + \mathcal{L}_u$, we have
    \begin{equation}
        \Big(\frac{\partial}{\partial t}+\mathcal{L}_u\Big)du^\flat
        = \,d\Big(\frac{1}{D}\frac{\partial U}{\partial T}\Big)\wedge dT.
    \end{equation}
    The advected scalar satisfies $(\partial_t+\mathcal{L}_u)\,dT
    = d\big((\partial_t+\mathcal{L}_u)T\big) = 0$, hence
    \begin{equation}
        \Big(\frac{\partial}{\partial t}+\mathcal{L}_u\Big)(\star_D q)
        = \Big(\frac{\partial}{\partial t}+\mathcal{L}_u\Big)\big(du^\flat\wedge dT\big)
        = \,d\Big(\frac{1}{D}\frac{\partial U}{\partial T}\Big)\wedge dT\wedge dT = 0.
    \end{equation}
    From the evolution of the
    potential vorticity $3$-form $\star_D q = q\,D\mu$ and the continuity
    equation, the potential vorticity scalar is advected.
\end{proof}

\begin{proposition}\label{prop:ErtelCasimir}
    The integral quantities 
    \begin{equation}
        C = \int_M\Phi(q)D\mu 
    \end{equation}
    with $q$ the Ertel potential vorticity are conserved quantities. 
\end{proposition}

\begin{proof}
    \begin{equation}
        \begin{aligned}
            \frac{d}{dt}C &= \frac{d}{dt}\int_M\Phi(q)D\mu\\
            &= \int_M \Phi'(q)\Big(\frac{\partial}{\partial t}q\Big)D\mu + \Phi(q)\frac{\partial}{\partial t}(D\mu)\\
            &= -\int_M \Phi'(q)(\mathcal{L}_uq)D\mu + \Phi(q)\mathcal{L}_u (D\mu) \\
            &= -\int_M \mathcal{L}_u\big(\Phi(q)D\mu\big)\\
            &= -\int_M d\big(\iota_u(\Phi(q)D\mu)\big) = 0,
        \end{aligned}
    \end{equation}
    by Stokes' theorem on a manifold without boundary.
\end{proof}

The proofs above illustrate why the presence of an additional advected field is essential: if no evolving scalar (or, more generally, 1-form field) is present, then the only natural 3-form one can build from the dynamical variables is
\begin{equation}
u^\flat\wedge du^\flat,
\end{equation}
the helicity 3-form, from which one can obtain the helicity scalar field. For Stokes' theorem to apply in the form above, we require a top form. Consequently, when no scalar field $T$ is advected, only constant multiples of helicity scalar field lie in the kernel of the associated Lie--Poisson bracket. By contrast, when an advected scalar field is included, any sufficiently smooth function of the Ertel potential vorticity $q$ yields a Casimir invariant. From a geometric perspective, the existence of nontrivial Casimirs is a consequence of having enough advected structure to construct Lie–transported top-degree forms constructed from the dynamical variables.

We now apply the energy-dissipation framework to the three-dimensional fluid system. We have established a Lie--Poisson structure on $\mathfrak{S}^*$ with the bracket $\{\,\cdot\,,\,\cdot\,\}_\mathfrak{S}$ given in \eqref{eq:lpbracket3d} and a Hamiltonian $\mathcal{H}$ given in \eqref{eq:ham3d}. To construct the symmetric bracket, we must first equip the configuration space with a Riemannian metric and select an appropriate functional $\mathcal{S}$ from the family of Casimir invariants. While the approach presented in this work is different from the selective decay theory presented in \cite{gay2013selective, gay2014geometric}, it produces similar dissipative terms.

\subsection{Irreversible incompressible fluids}
We now introduce the irreversible dynamics for the incompressible fluids in three dimensions. We seek to dissipate a Casimir invariant while conserving the energy $\mathcal{H}$. In 3D, the Casimir invariants are functionals of the Ertel potential vorticity $q$. Making the simplifying assumption $D=1$, the potential vorticity defined earlier 
\eqref{eq:potentialvorticity} becomes 
$q = \star(du^\flat \wedge dT)$. The Casimir invariants are of the form
\begin{equation}
\mathcal{S}(u, T) = \int_M \Phi(q) \mu,
\end{equation}
for arbitrary scalar functions $\Phi$.

\subsubsection{Potential Enstrophy}
Analogous to the 2D case, we select the potential enstrophy as the Casimir to define the entropy functional. Let $\Phi(q) = \frac{1}{2}q^2$. The functional is
\begin{equation}
\mathcal{S}_2(u, T) = \frac{1}{2\,\mathrm{Pe}}\int_M q^2 \mu,\end{equation}
where $\mathrm{Pe}=\frac{UL}{\kappa}$ is the P\'eclet number, which relates the typical velocity scale $U$ and length scale $L$ to the thermal diffusivity $\kappa$. The choice of the P\'eclet number as the relevant dimensionless parameter can be understood through the definition of potential vorticity, $q = \star(du^\flat \wedge dT)$, see \eqref{eq:potentialvorticity}. The vorticity $du^\flat$ describes the flow kinematics, whose diffusive scale is traditionally associated with the Reynolds number $\mathrm{Re} = UL/\nu$. The temperature gradient $dT$ introduces the thermodynamics of the active scalar, whose relation to momentum diffusion is governed by the Prandtl number $\mathrm{Pr} = \nu/\kappa$. Since potential vorticity couples these two fields, the dissipation coefficient is naturally governed by the product of these numbers, which yields the P\'eclet number: $\mathrm{Pe} = \mathrm{Re} \cdot \mathrm{Pr}$.

To define the symmetric bracket, we compute the variational derivatives of $\mathcal{S}_2$ with respect to $u$ and $T$. Using the definition of $q$, the variation is $\delta(q\mu) = d(\delta u^\flat)\wedge dT + du^\flat \wedge d(\delta T)$. Substituting this into the integral and integrating by parts yields
\begin{equation}
    \begin{aligned}
        \delta \mathcal{S}_2 &= \frac{1}{\mathrm{Pe}}\int_M q\,\delta (q\mu)\\
        &= \frac{1}{\mathrm{Pe}}\int_M q \Big( d(\delta u^\flat)\wedge dT + du^\flat \wedge d(\delta T) \Big) \\
        &= - \frac{1}{\mathrm{Pe}}\int_M \delta u^\flat \wedge (dT \wedge dq) - \frac{1}{\mathrm{Pe}} \int_M \delta T (dq \wedge du^\flat).
    \end{aligned}
\end{equation}
We can express the 2-form $dT \wedge dq$ in terms of a vector field via the metric. A natural choice is the metric induced by the kinetic energy of the fluid. In the general compressible case, this metric is mass-weighted. However, in the incompressible limit, the density $D$ becomes a fixed parameter $D_0$ (which we set to $1$ for simplicity), and the momentum density $1$-form $m$ is identified with the velocity field $u \in \mathfrak{X}_{\mathrm{div}}(M)$.

Restricting the kinetic energy metric to the subspace where $D=1$ yields the following $L^2$-type metric on the product space $\mathfrak{X}_{\mathrm{div}}(M) \times C^\infty(M)$
\begin{equation}\label{eq:metric3d}
    \langle (u_1, T_1), (u_2, T_2) \rangle_\mathfrak{S} = \int_M \Big(\langle u_1, u_2\rangle + T_1 T_2\Big) \mu.
\end{equation}
This metric identifies the dual space with the space itself, allowing us to define gradients. The gradients of the entropy functional are
\begin{equation}\label{eq:gradS3d}
    \frac{\delta \mathcal{S}_2}{\delta u} = - \frac{1}{\mathrm{Pe}}(\star(dT \wedge dq))^\sharp, \qquad \frac{\delta \mathcal{S}_2}{\delta T} = - \frac{1}{\mathrm{Pe}}\star(dq \wedge du^\flat).
\end{equation}
In terms of vector calculus, the term $\star(dT \wedge dq)$ corresponds to the vector product $\nabla T \times \nabla q$, and $\star(dq \wedge du^\flat)$ corresponds to the dot product $\nabla q \cdot \omega$, where $\omega$ is the vorticity vector. These terms are closely related to the selective decay theories presented in \cite{gay2013selective, gay2014geometric}. The irreversible equations of motion are given by
\begin{equation}\label{eq:navierstokes3D_PE}
    \begin{aligned}
        \frac{\partial}{\partial t} u^\flat + \mathcal{L}_u u^\flat &= -dp + \frac{1}{\mathrm{Pe}} \Big( \star(dT\wedge dq) - c(q,T) u^\flat \Big), \\
        \frac{\partial}{\partial t} T + \mathcal{L}_u T &=  \frac{1}{\mathrm{Pe}} \Big( \star(dq \wedge du^\flat) - c(q,T) U'(T) \Big),\\
        d^* u^\flat &= 0,
    \end{aligned}
\end{equation}
where $q = \star(du^\flat \wedge dT)$ is the potential vorticity and $c(q,T)$ is the nonlocal energy-conserving multiplier
\begin{equation}
    c(q,T) = \frac{\langle \star(dT \wedge dq), u^\flat \rangle_{L^2} + \langle \star(dq \wedge du^\flat), U'(T) \rangle_{L^2}}{|u^\flat|^2_{L^2} + |U'(T)|^2_{L^2}}.
\end{equation}
This system conserves total energy $\mathcal{H}$ and monotonically dissipates potential enstrophy $\mathcal{S}_2$. The dissipation is highly nonlinear and anisotropic. It acts specifically on the alignment between the gradients of temperature and potential vorticity. Unlike the 2D case where enstrophy dissipation leads to a Laplacian-like term, in 3D, minimizing potential enstrophy creates a driving force that attempts to align $\nabla T$ and $\nabla q$ (to vanish the wedge product) or smooth out gradients of $q$ relative to the flow.

\subsubsection{General potential vorticity moments}
We can generalise to arbitrary moments of the potential vorticity, $\mathcal{S}_p = \frac{1}{p}\int_M q^p \mu$. The gradients generalise straightforwardly
\begin{equation}
    \frac{\delta \mathcal{S}_p}{\delta u} = -(\star(dT \wedge dq^{p-1}))^\sharp, \qquad \frac{\delta \mathcal{S}_p}{\delta T} = -\star(dq^{p-1} \wedge du^\flat).
\end{equation}
Using these gradients in the symmetric bracket provides a family of selective decay models. For $p \gg 2$, the dynamics will selectively target extremal values of potential vorticity, effectively acting as a limiter on $q$ while leaving the bulk of the flow affected only by the nonlocal energy-restoration term $c(q,T)$.

\subsection{Irreversible compressible fluids}
The Ertel potential vorticity also defines the Casimirs for compressible fluids in three dimensions. We take the metric to be derived from the kinetic energy of the fluid. 
The state variables are the momentum-density coefficient $m \coloneqq Du^\flat$
(so that $m\otimes\mu\in\mathfrak{X}^*(M)$ is the momentum density itself),
the mass density $D\mu\in\mathrm{Dens}_\mu(M)$ (with coefficient function
$D$, Remark~\ref{rem:coefficient-convention}), and the advected scalar
field $T\in C^\infty(M)$.

We define the (degenerate) Riemannian metric on variations $(\delta m, \delta D, \delta T)$ as:
\begin{equation}\label{eq:kinetic_metric}
    \langle (\delta m_1, \delta T_1), (\delta m_2, \delta T_2) \rangle_\mathfrak{S} = \int_M \left( \frac{1}{D} \langle \delta m_1, \delta m_2 \rangle + D \delta T_1 \delta T_2 \right) \mu.
\end{equation}
Note that we omit the $\delta D$ term in the metric definition, effectively treating the metric as degenerate in the density direction to enforce zero mass diffusion. This is geometrically justified because the kinetic energy functional $\frac{1}{2}\int \|u\|^2 D \mu$ defines a norm on velocity fields with respect to the mass form $D\mu$, but acts linearly on the density itself. Consequently, the kinetic metric is naturally degenerate in the density direction, consistent with the physical assumption of vanishing mass diffusion.

Let $\Phi(q)=\frac{1}{2}q^2$, where the scalar potential vorticity is defined as $q=\frac{1}{D}\star(du^\flat\wedge dT)$, so that the potential enstrophy is given by 
\begin{equation}
    \mathcal{S}_2(m,D,T)=\frac{1}{2\,\mathrm{Pe}}\int_M q^2 D\mu,
\end{equation}
where $\mathrm{Pe}$ is the Peclet number. To define the symmetric bracket, we compute the variational derivatives of $\mathcal{S}_2$. The variation of the integral involves terms with respect to $m$, $T$, and $D$. We focus on the momentum and scalar variations. Using the identity $q \,D \mu = du^\flat \wedge dT$, we observe that the variation of the functional related to $m$ and $T$ is
\begin{equation}
    \begin{aligned}
        \delta \mathcal{S}_2 &= \frac{1}{\mathrm{Pe}}\int_M q\,\delta (du^\flat \wedge dT)\\
        &= \frac{1}{\mathrm{Pe}}\int_M q \Big( d(\delta u^\flat)\wedge dT + du^\flat \wedge d(\delta T) \Big) \\
        &= - \frac{1}{\mathrm{Pe}}\int_M \delta u^\flat \wedge (dT \wedge dq) - \frac{1}{\mathrm{Pe}} \int_M \delta T (dq \wedge du^\flat).
    \end{aligned}
\end{equation}
Note that while there is also a variation with respect to density $\frac{\delta \mathcal{S}_2}{\delta D} = \frac{1}{2\,\mathrm{Pe}}q^2$, the metric is degenerate in the density direction (i.e., no mass diffusion), so this term does not contribute to the dynamics. 

We can express the 2-form $dT \wedge dq$ in terms of a vector field via the metric. The gradients are
\begin{equation}\label{eq:gradS3d_comp}
    \frac{\delta \mathcal{S}_2}{\delta m} = - \frac{1}{\mathrm{Pe}}\star(dT \wedge dq), \qquad \frac{\delta \mathcal{S}_2}{\delta T} = - \frac{1}{\mathrm{Pe}}\star(dq \wedge du^\flat).
\end{equation}
The term $\star(dT \wedge dq)$ corresponds to the vector product $\nabla T \times \nabla q$, and $\star(dq \wedge du^\flat)$ corresponds to the dot product $\nabla q \cdot \omega$ in case the equations are presented in vector calculus form, similar to the incompressible case. The equations of motion are given by
\begin{equation}\label{eq:compressible_dissipation}
    \begin{aligned}
        \frac{\partial}{\partial t} (m\otimes\mu) + \mathcal{L}_u (m\otimes\mu) 
        &= \Big(- D\,d\Big(\frac{\partial U}{\partial D} -\frac12\langle u^\flat,u^\flat\rangle\Big) 
        +\frac{\partial U}{\partial T}\,dT + \frac{1}{\mathrm{Pe}} \Big( \star(dT \wedge dq) - c(q,D,T) m \Big)\Big)\otimes\mu, \\
        \frac{\partial}{\partial t} T + \mathcal{L}_u T &= \frac{1}{\mathrm{Pe}} \frac{1}{D} \Big( \star(dq \wedge du^\flat) - c(q,D,T)  \frac{\partial U}{\partial T} \Big), \\
    \frac{\partial}{\partial t} (D\mu) + \mathcal{L}_u (D\mu) &= 0.
    \end{aligned}
\end{equation}
Note the factor $1/D$ in front of the scalar equation but not the momentum equation: this follows from the mass metric \eqref{eq:kinetic_metric}, whose $D^{-1}$ weight on $\delta m$ cancels the $D^{-1}$ arising from $u^\flat=m/D$ in \eqref{eq:gradS3d_comp}, while its $D$ weight on $\delta T$ must instead be divided out to recover the Riesz representative.

The factor $c(q,D,T)$ is the nonlocal energy-conserving multiplier. While the equations are structurally similar to the incompressible equations, there is a fundamental difference. The incompressible equations are always nonlocal due to the pressure being solved via a Poisson equation. The usual compressible fluid equations are local in nature, but after introducing the irreversible terms, in the compressible case we now also obtain nonlocality. In other words, the global conservation of energy requires action at a distance. To be consistent with the kinetic metric geometry, the norms and pairings must be mass-weighted:
\begin{equation}
c(q,D,T) = \frac{\langle \star(dT \wedge dq), u^\flat \rangle_{L^2} + \langle \star(dq \wedge du^\flat), D^{-1} U_T \rangle_{L^2}}{\int_M D |u^\flat|^2 \mu + \int_M D^{-1} |U_T|^2 \mu},
\end{equation}
where $U_T = \partial U/\partial T$.
We can rewrite the momentum equation in \eqref{eq:compressible_dissipation} in terms of the velocity 1-form $u^\flat$. Since $m = D u^\flat$, and because the continuity equation holds, the $D$ factors effectively divide out, as in \eqref{eq:compressible-velocity}, leading to
\begin{equation}\label{eq:velocity_dissipation_comp}
    \frac{\partial}{\partial t} u^\flat + \mathcal{L}_u u^\flat = -d\left(\frac{\partial U}{\partial D}
    -\frac12\langle u^\flat,u^\flat\rangle\right)
    + \frac{1}{D}\frac{\partial U}{\partial T}\, dT + \frac{1}{\mathrm{Pe}} \frac{1}{D} \Big( \star(dT \wedge dq) - c(q,D, T) Du^\flat \Big).
\end{equation}
Notice that in the velocity formulation, the dissipation term is scaled by $1/D$. The dissipative force is a force density, so the resulting acceleration is inversely proportional to mass density. In regions of high density, the fluid is resisting entropic rearrangement, whereas in regions of low density, it is highly susceptible to it. 

This system conserves total energy $\mathcal{H}$ and monotonically dissipates potential enstrophy $\mathcal{S}_2$. Both formulations carry the
same kinetic contribution $-\tfrac12\langle u^\flat,u^\flat\rangle$ inside
the gradient; the difference is that the Lagrange multiplier $p$ enforcing
incompressibility is replaced here by the thermodynamic quantity
$\partial U/\partial D$, which is determined by the internal energy rather
than by a constraint. The force on the scalar field involves the thermodynamic dual $\partial U/\partial T$. The dissipative operators, however, retain the same geometric form as in the incompressible case, acting to align the gradients of temperature and potential vorticity.

\subsection{Helicity}\label{ssect:helicity}
Consider now a purely incompressible fluid on $M$ without any advected scalar quantities. The state is described solely by the velocity field $u \in \mathfrak{X}_{\mathrm{div}}(M)$ (or equivalently $u^\flat$). In this restricted setting, the Casimirs defined as functions of potential vorticity do not exist. The only remaining Casimir invariant is the helicity, which measures the topological linking of vortex lines.

Let the Hamiltonian be the kinetic energy $\mathcal{H}(u) = \frac{1}{2}\int_M \langle u, u\rangle \mu$. We choose the entropy functional $\mathcal{S}_{hel}$ to be the total helicity
\begin{equation}
\mathcal{S}_{hel}(u) = \frac{1}{2}\int_M u^\flat \wedge du^\flat = \frac{1}{2}\int_M \langle u, \omega \rangle \mu,\end{equation}
where $\omega = (\star du^\flat)^\sharp$ is the vorticity vector field. Unlike potential enstrophy, helicity is not positive definite. It can take positive or negative values depending on the chirality of the flow. To construct the irreversible dynamics associated with this choice of entropy, we calculate the variational derivative of the helicity with respect to the velocity.
\begin{align}
    \delta \mathcal{S}_{hel} &= \frac{1}{2}\int_M \left( \delta u^\flat \wedge du^\flat + u^\flat \wedge d(\delta u^\flat) \right) \\
    &= \frac{1}{2}\int_M \left( \delta u^\flat \wedge du^\flat + \left(du^\flat \wedge \delta u^\flat - d(u^\flat\wedge\delta u^\flat) \right)\right)\\
    &= \int_M \delta u^\flat \wedge du^\flat - \frac{1}{2} \int_M d(u^\flat\wedge\delta u^\flat).
\end{align}
The second term disappears by Stokes' theorem, since $M$ has no boundary, hence
\begin{equation}
    \delta \mathcal{S}_{hel} = \int_M \delta u^\flat \wedge du^\flat.
\end{equation}
Thus, the gradient of helicity with respect to the $L^2$ metric is simply the vorticity field
\begin{equation}
    \frac{\delta \mathcal{S}_{hel}}{\delta u} = (\star du^\flat)^\sharp = \omega.
\end{equation}
Note that in the two-dimensional computation for the variation of the enstrophy, we obtained \eqref{eq:varderivativeenstrophy}, which involves an extra derivative. This is to be expected, since the enstrophy in 2D is the square of the vorticity, whereas the helicity is the cross-product of velocity with vorticity. The latter involves one fewer derivative. This means that we will not obtain the Laplacian of the velocity field as the operator responsible for the dissipation.

The gradient of the energy is $\frac{\delta \mathcal{H}}{\delta u} = u$. The irreversible dynamics are constructed by projecting the helicity gradient onto the subspace orthogonal to the energy gradient. The projector $P^\perp_\mathcal{H}$ applied to $\omega$ is
\begin{equation}
P^\perp_\mathcal{H}(\omega) = \omega - \frac{\langle \omega, u \rangle_{L^2}}{|u|^2_{L^2}} u = \omega - \frac{\mathcal{S}_{hel}}{\mathcal{H}} u.
\end{equation}
The ratio $\mathcal{S}_{hel}/\mathcal{H}$ acts as a global geometric parameter characterizing the current ``helical density" of the flow relative to its energy. The resulting irreversible equations of motion, \eqref{eq:beltrami_relaxation} below, preserve energy while monotonically changing helicity:
\begin{equation}\label{eq:beltrami_relaxation}
\begin{aligned}
    \frac{\partial}{\partial t} u^\flat + \mathcal{L}_u u^\flat &= -dp - \omega^\flat +\frac{\mathcal{S}_{hel}}{\mathcal{H}} u^\flat, \\
    d^* u^\flat &= 0.
\end{aligned}
\end{equation}
Instead of the Laplacian, we now obtain a single derivative of the velocity, i.e. a term algebraic in the vorticity. This means that we do not obtain smoothing in the traditional sense of damping high frequencies. In fact, since the curl is a self-adjoint operator with eigenvalues unbounded in both directions, eigenmodes with eigenvalues tending to negative infinity grow exponentially, making the problem not well-posed in the sense of Hadamard. 

Nevertheless, we can make the following formal analogue to the two-dimensional case. A steady state of the irreversible operator occurs when $\omega^\flat= \frac{\mathcal{S}_{hel}}{\mathcal{H}}u^\flat$, which is the definition of a Beltrami field. Beltrami fields are known to be stationary solutions of the 3D Euler equations, and thus, similar to the two-dimensional case described in Theorem \ref{thm:thermodynamic_equilibrium}, the thermodynamic equilibrium characterized by the vanishing of the symmetric-bracket contribution, also yields a stationary solution for the dissipation-free equations. 

\section{Vector calculus descriptions}
In this section, we present the equations of motion for irreversible dynamics in terms of vector calculus. Suppose that the domain is $\mathbb{T}^2$ or $\mathbb{T}^3$, the two-torus or the three-torus, respectively. This is a flat and compact domain that we equip with the Euclidean metric. These are not simply-connected, so we have to impose mean-zero fields. Let $\mu = dx \coloneqq dx^1\wedge\cdots\wedge dx^n$,
the standard volume form of the Euclidean metric.
With this choice of $\mu$, the coefficient functions of
Remark~\ref{rem:coefficient-convention} coincide with the fields of
classical fluid dynamics: 
 $D$ equals the
classical mass density function $\rho$,
the mass form is $D \, \mu = \rho\,dx$. The momentum-density
coefficient $m = Du^\flat$ becomes the classical momentum density
$\rho\mathbf{u}$ under the Euclidean identification of $1$-forms with
vector fields.
The divergence operator
$\operatorname{div}_\mu$ becomes the usual
vector calculus divergence, and $\star$, $\sharp$ and $\flat$ reduce to the familiar
identifications of vector calculus.
The irreversible equations derived in Section \ref{sec:twodimensions} and \ref{sec:threedimensions} can then be expressed in terms of standard vector calculus operations, as follow.

\paragraph{Two-dimensional incompressible fluid.} 
The enstrophy-minimising irreversible dynamics in equation \eqref{eq:navierstokesgeneral} on $\mathbb{T}^2$ is 
given by 
\begin{equation}
    \begin{aligned}
        \frac{\partial}{\partial t}\mathbf{u} + (\mathbf{u}\cdot\nabla)\mathbf{u} &= -\nabla p + \frac{1}{\mathrm{Re}}\left(\Delta\mathbf{u}+\frac{\|\omega\|^2_{L^2}}{\|\mathbf{u}\|_{L^2}^2}\mathbf{u}\right),\\
        \nabla\cdot\mathbf{u}&=0,
    \end{aligned}
\end{equation}
where the scalar vorticity $\omega$ is defined as
\begin{equation}
    \omega=\nabla^\perp\cdot\mathbf{u}.
\end{equation}
The perpendicular gradient $\nabla^\perp$ is orthogonal in the Euclidean inner product to the gradient $\nabla$. These are the Navier-Stokes equations for an incompressible fluid in two dimensions with an additional nonlocal term that guarantees global energy conservation.

\paragraph{Three-dimensional incompressible fluid.}
Assuming that we are dealing with a three-dimensional incompressible fluid that is advecting a scalar field $T$, the potential enstrophy-minimising irreversible dynamics on $\mathbb{T}^3$ is given by
\begin{equation}
    \begin{aligned}
        \frac{\partial}{\partial t}\mathbf{u}+(\mathbf{u}\cdot\nabla)\mathbf{u} &= -\nabla p + \frac{1}{\mathrm{Pe}}\Big(\nabla T\times\nabla q - c(q,T)\mathbf{u}\Big),\\
        \frac{\partial}{\partial t}T+(\mathbf{u}\cdot\nabla)T &= \frac{1}{\mathrm{Pe}}\Big(\nabla q\cdot\boldsymbol{\omega} - c(q,T)U'(T)\Big),\\
        \nabla\cdot\mathbf{u} &= 0,
    \end{aligned}
\end{equation}
where the vorticity field $\boldsymbol{\omega}$, the potential vorticity scalar $q$ and the nonlocal multiplier $c(q,T)$ are given by
\begin{equation}\label{eq:symbols}
    \begin{aligned}
        \boldsymbol{\omega} &= \nabla\times\mathbf{u},\\
        q &= \boldsymbol\omega\cdot\nabla T,\\
        c(q,T) &= \frac{\int_{\mathbb{T}^3} (\nabla T\times\nabla q)\cdot\mathbf{u} + (\nabla q\cdot\boldsymbol{\omega})U'(T)\,dx}{\int_{\mathbb{T}^3}|\mathbf{u}|^2 + (U'(T))^2\,dx}.
    \end{aligned}
\end{equation}
The function $U(T)$ describes the internal energy of the fluid. Since the potential vorticity is linear in the vorticity, gradients of the potential vorticity involve the vector Laplacian of the velocity field. Gradients of the scalar field $T$ are also present, hence the diffusion is nonlinear and anisotropic.

If the three-dimensional incompressible fluid is not advecting a scalar field $T$, then the helicity-extremising irreversible dynamics on $\mathbb{T}^3$ is given by
\begin{equation}
    \begin{aligned}
        \frac{\partial}{\partial t}\mathbf{u} + (\mathbf{u}\cdot\nabla)\mathbf{u} &= -\nabla p -\boldsymbol\omega + \left(\frac{\int_{\mathbb{T}^3}\mathbf{u}\cdot\boldsymbol{\omega}\,dx}{\int_{\mathbb{T}^3}|\mathbf{u}|^2\,dx}\right)\mathbf{u},\\
        \nabla\cdot\mathbf{u}&=0,
    \end{aligned}
\end{equation}
where the vorticity field $\boldsymbol{\omega}$ is defined as in \eqref{eq:symbols}. In this case, the kinetic energy gives a complete description of the interactions inside the fluid and there is no internal energy. Note that there is no Laplacian present, the helicity-extremising irreversible dynamics involves only a single derivative of the velocity field. 

\paragraph{Three-dimensional compressible fluid.} 
Assuming that a three-dimensional compressible fluid is advecting a scalar field $T$, the potential enstrophy-minimising irreversible dynamics on $\mathbb{T}^3$ is given by
\begin{equation}
    \begin{aligned}
        \frac{\partial}{\partial t}\mathbf{u}+(\mathbf{u}\cdot\nabla)\mathbf{u} &= -\nabla\frac{\partial U}{\partial \rho} + \frac{1}{\rho}\frac{\partial U}{\partial T}\nabla T  + \frac{1}{\mathrm{Pe}}\frac{1}{\rho}\Big(\nabla T\times\nabla q - c(q,\rho,T)\rho\mathbf{u}\Big),\\
        \frac{\partial}{\partial t}T+(\mathbf{u}\cdot\nabla)T &= \frac{1}{\mathrm{Pe}}\frac{1}{\rho}\Big(\nabla q\cdot\boldsymbol{\omega} - c(q,\rho,T)\frac{\partial U}{\partial T}\Big),\\
        \frac{\partial}{\partial t}\rho + \nabla\cdot(\rho\mathbf{u}) &= 0,
    \end{aligned}
\end{equation}
where
\begin{equation}
    \begin{aligned}
        \boldsymbol{\omega} &= \nabla\times\mathbf{u},\\
        q &= \frac{\boldsymbol\omega\cdot\nabla T}{\rho},\\
        c(q,\rho,T) &= \frac{\int_{\mathbb{T}^3} (\nabla T\times\nabla q)\cdot\mathbf{u} + \frac{1}{\rho}(\nabla q\cdot\boldsymbol{\omega})\frac{\partial U}{\partial T}\,dx}{\int_{\mathbb{T}^3}\rho|\mathbf{u}|^2 + \frac{1}{\rho}(\frac{\partial U}{\partial T})^2\,dx}
    \end{aligned}
\end{equation}

The function $U(\rho,T)$ describes the internal energy per unit volume of
the fluid, a function of both the mass density coefficient $\rho$ and the
advected scalar $T$.

\section{Discussion and conclusion}

We work with a constructive definition of a class of irreversible 
(dissipative) systems on 
Poisson manifolds with a Riemannian metric, 
within the paradigm of metriplectic systems.
By design, these systems conserve energy and optionally other integrals of motion,
while one or more scalar functions, called generalised entropies, evolve 
monotonically. In contrast to the original metriplectic paradigm,
our generalised entropies need not be Casimirs, 
but only integrals of motion.
This means that the dissipated functional $\mathcal S$ need only
Poisson-commute with the energy, $\{\mathcal S,\mathcal H\}=0$, and need not be a
Casimir; this is what allows conserved quantities such as functions of the angular momentum, to
serve as $\mathcal S$. For instance, in many finite-dimensional cases, such as central force problems, one works with a phase space that is a symplectic manifold before passing to a Poisson or symplectic reduced space. Non-Casimir generalized entropies are especially useful in this case, since the Casimirs on a symplectic manifold are trivial.

In addition, our symmetric bracket is \textrm{constructed geometrically} from 
a given Riemannian metric with no special properties. This allows us to apply the framework on orientable, smooth Riemannian manifolds without boundary in dimensions two or three and express the resulting equations in a coordinate free way. The same formulation can be applied, for example, to the sphere relevant to geophysical flows and on the (twisted) tori relevant
to plasma confinement.

In the two-dimensional setting, our framework produces an energy-conserving version of the Navier-Stokes equations from the Euler equations for ideal fluids, when the entropy is given by the enstrophy. As shown in Theorem \ref{thm:thermodynamic_equilibrium}, the states with zero symmetric bracket contribution, i.e. zero enstrophy dissipation, correspond to the stream function being a Laplace–Beltrami eigenfunction, which, in turn, are stationary solutions of both the Euler and metriplectic equations. Moreover, Proposition \ref{prop:poincare} shows that the set of the enstrophy-minimizing states with fixed kinetic energy lies in the first eigenspace. While enstrophy decreases monotonically and converges, the convergence of the velocity field to an equilibrium remains an open question. Additionally, we also derive the resulting equations when the entropy is given by higher moments of the vorticity. We note that, in all these cases our framework naturally introduces a non-local term that restores the global energy balance. 

The three-dimensional case, however, is different from the two-dimensional one. As explained in Remark \ref{rem:3D-NS}, while we can use enstropy to obtain energy-conserving Navier-Stokes in three-dimensions from Euler equations, this is not a generalized entropy since it is not conserved by the Euler flow. Since the evolution for the enstrophy is not necessarily monotonic, from a physical point of view this is inconsistent with its interpretation as a thermodynamic entropy.

In the three-dimensional setting, for a purely incompressible fluid, the only Casimir is the helicity. When this is used as the entropy function, the resulting equations contain only a single derivative of the velocity instead of the Laplacian. The fundamental reason for this difference lies in the construction of a scalar vorticity variable. In the two dimensional setting, the vorticity is already a scalar. In three dimensions, the curl of the velocity field is vector-valued. We also note that in this case the condition for zero-helicity dissipation selects stationary solutions for the dissipation-free Euler equations.

To obtain a scalar that is materially transported, additional dynamical variables are required. These additional variables modify the dissipation and lead to highly anisotropic diffusion. These variables are introduced in the form of an additional scalar or a 1-form field, and yield a richer set of Casimirs in the form of functions of the potential vorticity. Using moments of the potential vorticity, particularly the potential enstrophy, as entropies, we derive equations of motion for three-dimensional compressible and incompressible fluids. It would be interesting to see whether this alternative to the three-dimensional Navier-Stokes equations behaves better with regards to global existence and uniqueness of strong solutions. This is outside of the scope of this work.

\bibliographystyle{plainnat}
\bibliography{biblio}
\end{document}